\documentclass[%
 reprint,
 superscriptaddress,
 amsmath,amssymb,
 aps,
 pra,
]{revtex4-2}

\usepackage{graphicx}
\usepackage{dcolumn}
\usepackage{bm}
\usepackage{xcolor}
\newcommand{\D}{\mathrm{d}}

\begin{document}

\preprint{APS/123-QED}

\title{Three-dimensional canonical quantum plasmonics for finite media: \\exact solution in terms of the classical Green tensor}

\author{Georgii Semin}
\affiliation{Laboratoire Interdisciplinaire Carnot de Bourgogne ICB UMR 6303, Université Bourgogne Europe, CNRS, F-21000 Dijon, France}
\author{Hans-Rudolf Jauslin} 
\affiliation{Laboratoire Interdisciplinaire Carnot de Bourgogne ICB UMR 6303, Université Bourgogne Europe, CNRS, F-21000 Dijon, France}
\author{St\'{e}phane Gu\'{e}rin}
\affiliation{Laboratoire Interdisciplinaire Carnot de Bourgogne ICB UMR 6303, Université Bourgogne Europe, CNRS, F-21000 Dijon, France}

\date{\today}

\begin{abstract}
This article presents a comprehensive three-dimensional canonical quantization to treat quantum plasmonics for finite metallic or dielectric media of arbitrary shape. We use a microscopic model for the dissipative and dispersive medium coupled with the electromagnetic field, which is justified by the fact that if one integrates the degrees of freedom of the medium, one obtains the macroscopic Maxwell equations. Its quantization features a Hamiltonian formulation having the form of two infinite harmonic oscillators characterized by a double continuum. The diagonalized Hamiltonian is quantized by the correspondence principle, introducing creation-annihilation operators in a bosonic Fock space. The diagonal quantum Hamiltonian is the sum of two terms corresponding to the two continua. The physical observables, like, e.g., the electric field, are also the sum of two terms corresponding to the two continua, one of which had been omitted in the literature geared for an infinite bulk medium. In a second step, we show that the electric field operator can by written as linear combinations of the creation-annihilation operators with coefficients that satisfy integral equations of Fredholm type. We show that the solution of these equations can be expressed in terms of the classical Green tensor of the medium satisfying the Sommerfeld radiation condition. Finally, we consider the Purcell effect for the spontaneous emission of an atom close to the medium. We show that through an exact compensation of some terms, the Purcell factor for the system with the double continuum is proportional to the imaginary part of the Green tensor, which defines the local density of states. This result has the same form as the one obtained in the literature for bulk systems that involve a single continuum and a small dissipative background extending to infinity, and can be seen as a justification of this approach.
\end{abstract}

\maketitle


\section{Introduction}

Quantum plasmonics studies the interaction between quantum electromagnetic fields and matter on a quantum scale, where plasmonic nanostructures are taken as the medium \cite{garrison2008quantum, novotny2012principles, Tame2013, Zhou2019, benisty2022introduction}. The coupling of the photons with plasmons in media leads to new quasi-particles called quantized plasmon polaritons (QPP). 

Two approaches, based on microscopic and macroscopic descriptions of a medium, have been widely used for the theoretical description of quantum plasmonics. As for the first one, Huttner and Barnett proposed a description with an infinite homogeneous medium represented by oscillators coupled to the electromagnetic field \cite{Huttner1991, Huttner1992a, Huttner1992b, Barnett1992, Hopfield1958}. They employed the technique introduced by Friedrichs \cite{friedrichs1948perturbation, friedrichs1965} and Fano \cite{fano1961effects} to diagonalize the Hamiltonian. This approach was later extended for systems with an inhomogeneous medium \cite{Wubs2001, Suttorp2004, Suttorp2004a, Wubs2004, bhat2006hamiltonian}. On the other hand, a phenomenological approach introduced by Gruner and Welsch was formulated in terms of a Langevin noise formalism \cite{Gruner1995, Gruner1996, Dung1998three, Scheel1998, Scheel2008, Vogel2006, Matloob1995, Matloob1997} and based on the Green tensor,  which was also adapted for an inhomogeneous medium \cite{Drezet2016, Drezet2017a, Drezet2017b}. 

Both approaches consider the medium as an infinite bulk. It was noticed \cite{DiStefano2000, Stefano2001} that if one tries to apply the formulas constructed for a bulk to a finite medium, there is a conceptual difficulty: If one takes the uncoupling limit on the bulk formulas, the electric field observable becomes zero. An idea to get around this difficulty was proposed, e.g. in \cite{buhmann2013dispersion, Drezet2017b, Hanson2021} consisting in adding a small dissipative background $\varepsilon_{bg}$ extending to infinity, which justifies the use of the bulk formulas, and at the end of the calculation taking the limit  $\varepsilon_{bg}\to 1$. It was remarked in \cite{Dorier2020} that this procedure might work for some physical quantities, but it is difficult to justify that it works for all physical observables. Several modified Langevin noise approaches were developed to resolve this issue by adding a free field to the medium-assisted field \cite{Drezet2017a, Na2023, Ciattoni2024, miano2025quantum}. 

In \cite{Dorier2019}, an approach of canonical quantization of the Hamiltonian of the system with a finite medium was formulated, in which the Hamiltonian diagonalization was performed using the Lippmann-Schwinger equations. The main result of that article is that the Hamiltonian for a finite medium has a double  continuous spectrum, as opposed to the bulk systems where there is just one continuous spectrum. It was proven that the Hamiltonian of the electromagnetic field coupled to the medium is unitarily equivalent to the uncoupled model, which has a double continuum structure. This approach was implemented in detail for a one-dimensional model of the electromagnetic field coupled with a homogeneous finite medium in \cite{Semin2024}. It was shown that the eigenfunctions and the electric field operators are composed of two terms that can be expressed exactly in terms of the classical Green function of the medium. With these results, the decay rate of an atom in or outside the medium was calculated. The result showed that the Purcell factor calculated with the double continuum approach coincides with the one obtained in the literature using the bulk formulas obtained by adding a dissipative background. This result had also been shown with numerical simulations in Ref. \cite{Na2023}.

In the present work, we extend to three dimensions the results obtained for one dimension in \cite{Semin2024}, where many of the results used the fact that the Green function for a homogeneous slab can be written explicitly in terms of elementary functions. The present results are more general in that they provide the canonical quantization and the exact diagonalization for a finite inhomogeneous media of arbitrary shape in three dimensions. 

The main results of the article can be summarized as follows:\\
(i) We show that the diagonal quantized Hamiltonian is the sum of two terms associated respectively through the uncoupling limit to the electromagnetic field and the medium:
\begin{align}
\label{HamCem}
\hat{H}=&\int \D^3k \sum_{\sigma, \zeta}\, \hbar \omega_{\kappa} \hat{C}_{\mathbf{k}, \sigma, \zeta}^{e\dagger} \hat{C}^e_{\mathbf{k}, \sigma, \zeta}\notag \\
&+ \int \D^3 x \int \D \nu \sum_j \, \hbar \nu \hat{C}_{\mathbf{x}, \nu, j}^{m\dagger} \hat{C}^m_{\mathbf{x}, \nu, j},
\end{align}
where the two families of bosonic creation-annihilation operators $\hat{C}_{\mathbf{k}, \sigma, \zeta}^{e\dagger}$, $\hat{C}_{\mathbf{k}, \sigma, \zeta}^{e}$ and $\hat{C}_{\mathbf{x}, \nu, j}^{m\dagger}$, $\hat{C}_{\mathbf{x}, \nu, j}^{m}$ are respectively related to the free electromagnetic and the medium excitations in the uncoupling limit with the wave vector $\mathbf{k}$, $\omega = c|\mathbf{k}|$, polarization $\sigma$, index of symmetry $\zeta$, position $\mathbf{x}$, oscillator frequency $\nu$, and the direction $j$. \\
(ii) From the principle of the canonical quantization based on two continua, we derive the explicit expression of the electric field observable, which is also the sum of two terms
\begin{align}
\hat{\mathbf{E}}(\mathbf{r}) &= \hat{\mathbf{E}}^e(\mathbf{r}) + \hat{\mathbf{E}}^m(\mathbf{r}), \\
\hat{\mathbf{E}}^e(\mathbf{r}) &= -\int \D\kappa \sqrt{\frac{\hbar}{2\varepsilon_0 \omega_\kappa }}\left[ \mathbf{e}_\kappa(\mathbf{r}) \hat{C}^e_\kappa + \text{H.c.}\right], \label{def:Ee} \\
 \hat{\mathbf{E}}^m(\mathbf{r}) &= -\int \D\mu \sqrt{\frac{\hbar}{2\varepsilon_0 \nu }}\left[ \mathbf{m}_\mu(\mathbf{r}) \hat{C}^m_\mu + \text{H.c.}\right], \label{def:Em}
\end{align}
and we construct integral equations for the $ \mathbf{e}$ and $ \mathbf{m}$ coefficients from the Lippmann-Schwinger equations
\eqref{integral-eq-e-hrj},  \eqref{eqn:m_int}. \\
(iii) One of the main results of this article is that the coefficients $ \mathbf{e}_\kappa(\mathbf{r})$  and $\mathbf{m}_\mu(\mathbf{r}) $ can be  written explicitly in terms of the classical Green tensor for the medium  $\bar{\bar{G}}_{m+}$ satisfying the Sommerfeld radiation condition 
\eqref{eqn:int_e_2}:
\begin{align}
	\mathbf{e}_\kappa(\mathbf{r}) =&  \omega_{\kappa}\mathbf{\varphi}_\kappa(\mathbf{r}) +  \frac{\omega^2_\kappa}{c^2} \int_V \D^3x\, \bar{\bar{G}}_{m+}(\mathbf{r}, \mathbf{x}, \omega_\kappa)\notag \\ 
	&\times[\varepsilon(\mathbf{x}, \omega_\kappa) - 1]\omega_{\kappa}\mathbf{\varphi}_\kappa(\mathbf{x}), 
\end{align}
where  $\mathbf{\varphi}_\kappa$ are the continuum eigenfunctions of $\nabla\times\nabla\times$,  and \eqref{eqn:m_green}:
\begin{equation}
	\mathbf{m}_{\mathbf{x}, \nu, j}(\mathbf{r}) = -\tilde{\alpha}(\mathbf{x}, \nu)\frac{\nu^2}{c^2}\bar{\bar{G}}_{m+}(\mathbf{r}, \mathbf{x}, \nu) \cdot \mathbf{n}_j,
\end{equation}
where $\mathbf{n}_j$ are unit vectors in direction $j$. \\
(iv) We also show that the eigenfunctions $\underline \psi_\kappa^e,  \underline \psi_\mu^m$, solutions of the Lippmann-Schwinger equations can be written explicitly in terms of integrals of the Green tensor  $\bar{\bar{G}}_{m+}$.\\
(v) With these results, one can, e.g., analyze the Purcell effect of an atom close to or inside the medium. For that, we use the following local density of states (LDOS) identity that relates the imaginary part of the Green tensor to the $\mathbf{e}_\kappa$ coefficients \eqref{LDOS-identity-hrj}: 
\begin{align}  \label{LDOS-identity-hrj-44}
  &  \mathrm{Im}[\bar{\bar{G}}_{m}(\mathbf{x}, \mathbf{y}, \omega_{\kappa})] = \frac{\pi c^2}{2 \omega^3}\sum_{\sigma, \zeta} \mathbf{e}_{\omega, \sigma, \zeta}(\mathbf{x}) \otimes \mathbf{e}^*_{\omega, \sigma, \zeta}(\mathbf{y}) \notag \\
    &+ \frac{\omega^2}{c^2} \int \D^3 z\, \varepsilon_i(\mathbf{z}, \omega_{\kappa}) \bar{\bar{G}}_{m}(\mathbf{x},\mathbf{z}, \omega_{\kappa}) \bar{\bar{G}}_{m}^*(\mathbf{z}, \mathbf{y}, \omega_{\kappa}).
\end{align}\\
(vi) The decay rate by spontaneous emission involves two contributions \eqref{Gae-term}, \eqref{Gam-term}
\begin{align}
\Gamma(\mathbf{r}_a, \omega_a) = \Gamma_e(\mathbf{r}_a, \omega_a) + \Gamma_m(\mathbf{r}_a, \omega_a),
\end{align} 
with
\begin{align}
&\Gamma_e(\mathbf{r}_a, \omega_a) \notag \\ 
&= \frac{\pi}{\hbar\omega\varepsilon_0}\int& \D\kappa\, \delta(\omega_\kappa - \omega_a)
 \mathbf{d}\cdot \mathbf{e}_{\kappa}(\mathbf{r}_a) \otimes \mathbf{e}^*_{\kappa}(\mathbf{r}_a)  \cdot \mathbf{d}
\label{Gae-term-00}
\end{align}
and
\begin{align} 
&\Gamma_m(\mathbf{r}_a, \omega_a) = \frac{2\omega_a^2}{\hbar\varepsilon_0c^2} \mathbf{d} \cdot \mathrm{Im}[\bar{\bar{G}}_{m+}(\mathbf{r}_a, \mathbf{r}_a, \omega)] \cdot \mathbf{d}\notag \\
& - \frac{\pi}{\hbar\omega\varepsilon_0} 
\int \D\kappa\, \delta(\omega-\omega_\kappa)\mathbf{d}\cdot \mathbf{e}_{\kappa}(\mathbf{r}_a) \otimes \mathbf{e}^*_{\kappa}(\mathbf{r}_a) \cdot \mathbf{d}.
\label{Gam-term-00}
\end{align}
We observe  that the $ \mathbf{e}_\kappa\otimes \mathbf{e}_\kappa^*$ term in \eqref{Gam-term-00} exactly compensates \eqref{Gae-term-00}, the contribution of the $\hat{\mathbf{E}}^e$ term to the decay rate, and thus we obtain \eqref{Ga_final}:
\begin{align} \label{Ga_final_00}
\Gamma(\mathbf{r}_a, \omega_a) = 
 \frac{2\omega_a^2}{\hbar\varepsilon_0c^2} \mathbf{d} \cdot \mathrm{Im}[\bar{\bar{G}}_{m+}(\mathbf{r}_a, \mathbf{r}_a, \omega)] \cdot \mathbf{d}
\end{align} 
which coincides with the formula reported in the literature \cite{novotny2012principles}, that was obtained using formulas for a bulk medium that do not contain the 
 $\hat{\mathbf{E}}^e$ contribution to the electric field nor the $ \mathbf{e}_\kappa(\mathbf{x}) \otimes \mathbf{e}^*_\kappa(\mathbf{y})$ term in the identity \eqref{LDOS-identity-hrj-44}.
The exact compensation of these two terms provides an explanation of the fact that the formulas for a bulk medium, that in principle cannot be applied to a finite medium, lead to the same result obtained with the formulas for a finite medium.This clarifies a question that  has been difficult to settle in the literature for a long time \cite{DiStefano2000, Stefano2001, buhmann2013dispersion, Drezet2016, Drezet2017a, Drezet2017b, Hanson2021, khanbekyan2005qed, khanbekyan2003input}.
 
The article is structured as follows: In Section \ref{sec:Model}, we describe the model and introduce its classical Hamiltonian. Then, we define the canonical variables and demonstrate the process of diagonalizing the Hamiltonian using the Lippmann-Schwinger equations, and we proceed to quantize the Hamiltonian. In Section \ref{sec:EM_field}, we define the electric field operator in terms of the creation-annihilation operators associated with the free electromagnetic field and the medium and the field coefficients and derive the integral equations for them. We show the relationship between the classical Green tensor of the medium and the field coefficients. In Section \ref{sec:Prop}, we demonstrate the relationship between the field coefficients and the eigenfunctions of the coupled Hamiltonian, we prove the Green tensor LDOS identity and its application for the Purcell factor, and we show that in the uncoupling limit, the model acts as a model of the free field.  Finally, in Section \ref{sec:Conclusion}, we summarize the results.

\section{Model}\label{sec:Model}

\subsection{Classical Hamiltonian}
\begin{figure}[h]
   \centering
   \includegraphics[width=0.5\textwidth]{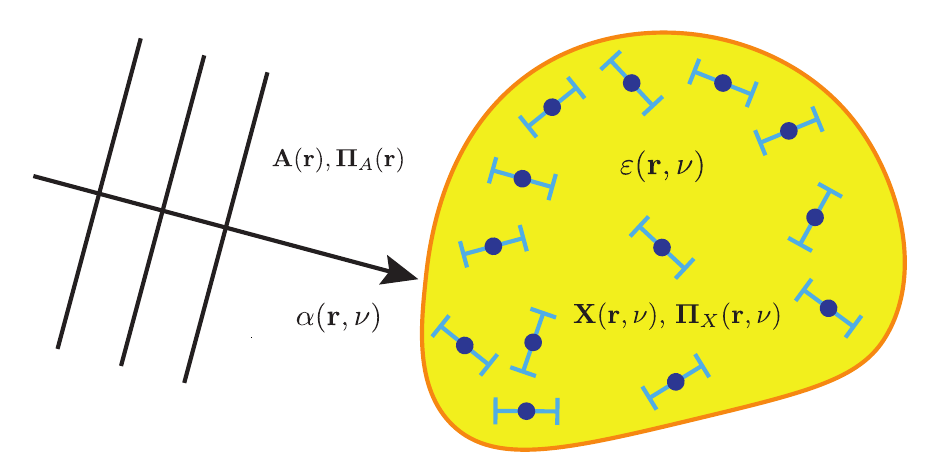}
   \caption{Scheme of the system of the electromagnetic field with vector potential $\mathbf{A}(\mathbf{r})$ and its conjugated momentum $\mathbf{\Pi}_{A}(\mathbf{r})$ and the finite inhomogeneous dielectric medium with canonical variables $\mathbf{X}$ and $\mathbf{\Pi}_X$ and the dielectric coefficient $\varepsilon(\mathbf{r}, \nu)$. The interaction between the field and the medium is defined by the coupling coefficient $\alpha(\mathbf{r}, \nu)$.}
   \label{fig:Model}
\end{figure}

We start by considering the model of the electromagnetic field and the finite-size inhomogeneous non-magnetic medium located in the volume $V$ (see Fig.\ref{fig:Model}). We consider the following classical Hamiltonian, compatible with the Maxwell equations \cite{Philbin2010},
\begin{equation}\label{eqn:Hamiltonian}
    H = H_{\mathrm{em}} + H_{\mathrm{med}} + H_{\mathrm{int}}
\end{equation}
with the Hamiltonian of the electromagnetic field
\begin{equation}
    H_{\mathrm{em}} = \frac{1}{2}\int_{\mathbb{R}^3} \D^3r \left[\frac{1}{\varepsilon_0} \mathbf{\Pi}^2_A(\mathbf{r}) - c^2\varepsilon_0 \mathbf{A}(\mathbf{r})\cdot \Delta\mathbf{A}(\mathbf{r})\right]
\end{equation}
expressed in terms of dielectric permittivity of vacuum $\varepsilon_0$, speed of light $c$, vector potential $\mathbf{A}$ and conjugated momentum $\mathbf{\Pi}_A$ expressed as
\begin{equation}\label{def:momentum_A}
    \mathbf{\Pi}_A(\mathbf{r}) = -\varepsilon_0 \mathbf{E}(\mathbf{r}) - \int_0^\infty \D\nu\, \alpha(\mathbf{r}, \nu)\mathbf{X}(\mathbf{r}, \nu),
\end{equation}
the free Hamiltonian of the medium oscillations
\begin{equation}
    H_\mathrm{{med}} = \frac{1}{2}\int_0^\infty \D \nu \int_V \D^3r \left[ \mathbf{\Pi}_X^2 + \nu^2 \mathbf{X}^2\right] 
\end{equation}
expressed in terms of position $\mathbf{X}(\mathbf{r}, \nu)$ and momentum $\mathbf{\Pi}_X(\mathbf{r}, \nu)$ of the oscillator with the frequency $\nu$ and position $\mathbf{r}$, and the interaction term
\begin{equation}
    H_{\mathrm{int}} = H_{c} + H_{s}.
\end{equation}
It consists of the cross-interaction term
\begin{equation}
    H_{c} = \frac{1}{\varepsilon_0} \int_{\mathbb{R}^3} \D^3r\, \mathbf{\Pi}_A \cdot \mathcal{P}^\perp \int_0^\infty \D \nu\, \alpha \mathbf{X}
\end{equation}
and self-interaction term
\begin{equation}
    H_{s} = \frac{1}{2\varepsilon_0} \int_V \D^3r \left[\int_0^\infty \D \nu\, \alpha \mathbf{X} \right]^2.
\end{equation}
Here, $\mathcal{P}^\perp$ is the projector on transverse fields, and $\alpha(\mathbf{r}, \nu)$ is the coupling function. We adopt the Coulomb gauge $\nabla \cdot \mathbf{A} = 0,$ $\nabla \cdot \mathbf{\Pi}_A = 0$, which leads to the transversality condition of the field, e.g. $\mathcal{P}^\perp \mathbf{\Pi}_A = \mathbf{\Pi}_A$.

The dielectric properties of the medium are given by the dielectric coefficient $\varepsilon(\mathbf{r}, \nu)$, which is 
\begin{equation}\label{def:diel_coef}
    \varepsilon(\mathbf{r}, \nu) = 
    \begin{cases}
        \varepsilon_{m}(\mathbf{r}, \nu) & \text{if } \mathbf{r} \in  V, \\
        1 & \text{if } \mathbf{r} \notin V.
    \end{cases}
\end{equation}
The coupling coefficient is determined by the imaginary part of the dielectric coefficient and takes the form 
\begin{equation}
    \alpha(\mathbf{r}, \nu) = \sqrt{\frac{2\varepsilon_0\nu}{\pi}\mathrm{Im}\,\varepsilon(\mathbf{r}, \nu)}.
\end{equation}
To preserve the causality principle, the dielectric coefficient must obey the Kramers-Kronig relation \cite{landau2013electrodynamics}
\begin{equation}\label{eqn:KK}
\int \D\nu \frac{\tilde{\alpha}^2(\mathbf{r}, \nu)}{\nu^2 - \lambda^2 - i0^+} = \varepsilon(\mathbf{r}, \lambda) - 1,
\end{equation}
with $\tilde\alpha:=\alpha/\sqrt{\varepsilon_0}$.

\subsection{Canonical variables of the electromagnetic field}\label{sec:canon_var}

We expand the fields $\mathbf{A}$ and $\mathbf{\Pi}_A$ in the transverse basis of real plane waves, which are eigenfunctions of the self-adjoint operator $\nabla \times \nabla \times$ on the Hilbert space $L_2(\mathbb{R}^3, \mathbb{C}^3, \mathrm{d}^3r)$:
\begin{subequations}
\begin{align}
	\mathbf{A}(\mathbf{r}) &= - \int \D\kappa\, \frac{1}{\sqrt{\varepsilon_0}\,\omega_\kappa} \mathbf{\Phi}_\kappa (\mathbf{r}) p_\kappa, \label{par:A} \\
	\mathbf{\Pi}_A(\mathbf{r}) &= \int \D \kappa\, \sqrt{\varepsilon_0}\,\omega_\kappa \mathbf{\Phi}_\kappa (\mathbf{r}) q_\kappa, \label{par:momentum_A}
\end{align}
\end{subequations}
with
\begin{equation}
    \nabla \times \nabla \times \mathbf{\Phi}_\kappa (\mathbf{r}) - \frac{\omega_\kappa^2}{c^2}\mathbf{\Phi}_\kappa (\mathbf{r}) = 0.
\end{equation}
Here, $\kappa$ is a multi-index containing the eigenvalue and some indices that label the degeneracy. The choice of a complete set of a continuum orthonormal basis is not unique. Here, we consider the real-valued set formed by linearly polarized plane waves 
\begin{equation} \label{def-Phi_kappa}
	\mathbf{\Phi}_\kappa (\mathbf{r}) = (2\pi)^{-3/2}\vec{\epsilon}_{\sigma_{\kappa}} 
	\begin{cases}
	\cos(\mathbf{k}\mathbf{r}) & \zeta_\kappa = c, \\
	\sin(\mathbf{k}\mathbf{r}) & \zeta_\kappa = s,
	\end{cases}
\end{equation}
where $\mathbf{k} \in \mathbb{R}^2\times[0, +\infty)$ is the wavevector and $\omega_\kappa = c|\mathbf{k}|$, $\vec{\epsilon}_{\sigma_{\kappa}}$ with $\sigma_\kappa = \pm$ are two arbitrary real unit vectors orthogonal to $\mathbf{k}$ and to each other. For this choice the multi-index is $\kappa = (\mathbf{k}, \sigma_\kappa, \zeta_\kappa)$ with the following integration rule 
\begin{equation}
\int \D \kappa = \int \D^3k \sum_{\sigma_\kappa = \pm} \sum_{\zeta_\kappa=c,s}.
\end{equation}
The orthonormality of the chosen set of eigenvectors is given by
\begin{align}
\int_{\mathbb{R}^3}\D^3r\,  \mathbf{\Phi}_\kappa (\mathbf{r}) \cdot \mathbf{\Phi}_{\kappa'} (\mathbf{r}) &= \delta(\kappa - \kappa') \notag \\
&= \delta(\mathbf{k} - \mathbf{k}')\delta_{\sigma_{\kappa}\sigma_{\kappa'}}\delta_{\zeta_{\kappa}\zeta_{\kappa'}},
\end{align}
and completeness in the space of transverse fields is given by
\begin{equation}
\int \D\kappa\, \mathbf{\Phi}_\kappa (\mathbf{r}) \otimes \mathbf{\Phi}_{\kappa} (\mathbf{r}') = \delta^\perp(\mathbf{r} - \mathbf{r}'),
\end{equation}
where $\delta^\perp(\mathbf{r} - \mathbf{r}') $ is the transverse delta-funciton. The reason for choosing a real basis $\{\mathbf{\Phi}_\kappa (\mathbf{r})\}$ is that (\ref{par:A}), (\ref{par:momentum_A}) are real canonical transformations.

\subsection{Hamiltonian in harmonic oscillator-like form}

The Hamiltonian in terms of the variables $(q, p, \mathbf{X}, \mathbf{\Pi}_X)$ takes the form $H = H_{EM} + H_{med} + H_{c} + H_{s}$, where 
\begin{subequations} \label{microscopic-Hamiltonian-hrj}
\begin{align}
    H_{EM} =& \frac{1}{2} \int \D\kappa \left[p^2_\kappa + \omega^2_\kappa q^2_\kappa\right], \\
    H_{med} =& \frac{1}{2} \int \D\mu \left[\mathbf{\Pi}^2_X + \nu^2 \mathbf{X}^2 \right],\\
    H_c =& \int \D\kappa\, \omega_\kappa\, q_\kappa \int_{\mathbb{R}^3} \D^3x\, \mathbf{\Phi}_\kappa(\mathbf{x}) \notag \\
    &\times \int_0^\infty \D\nu\, \tilde{\alpha}(\mathbf{x}, \nu) \mathbf{X}(\mathbf{x}, \nu),\\
    H_s =& \frac{1}{2} \int_V \D^3x \left[\int_0^\infty \D\nu\,\tilde{\alpha}(\mathbf{x}, \nu) \mathbf{X}(\mathbf{x}, \nu)\right]^2.
\end{align}
\end{subequations}

Introducing the following notation
\begin{equation}
\left(f | g \right) = 
\begin{cases}
\int \D x\, f(x)g(x) & \text{if } x \text{ is continuous}, \\
\sum_i f(x_i)g(x_i) & \text{if } x \text{ is discrete},
\end{cases}
\end{equation}
we rewrite the terms of the Hamiltonian into an integral-operator form
\begin{subequations}
\begin{align}
    H_{EM} &= \frac{1}{2}\left[ \left(p | p \right) + \left(q | \omega^2_\kappa q \right) \right], \\
    H_{med} &= \frac{1}{2}\left[ \left(\mathbf{\Pi}_X |\mathbf{\Pi}_X\right) + \left(\mathbf{X}| \nu^2 \mathbf{X}\right) \right], \\
    H_{c} &= \frac{1}{2}\left[\left( q | B_1 \mathbf{X} \right) +  \left(\mathbf{X}|B_2 q \right) \right], \\
    H_{s} &= \frac{1}{2}\left(X | A X\right),
\end{align}
\end{subequations}
with the integral operators
\begin{subequations}
\begin{align}
[B_1 \mathbf{X} ](\kappa) =& \int_V \D^3 x_\mu \bigg(\mathbf{\Phi}_\kappa(\mathbf{x}_\mu)  \cdot \int_0^\infty \D \nu_\mu\,  \notag \\
&\times \omega_\kappa \tilde{\alpha}(\mathbf{x}_\mu, \nu_\mu)\mathbf{X}(\mathbf{x}_\mu, \nu_\mu)\bigg), \label{def:B1}\\
[B_2 q](\mu) =& \tilde{\alpha}(\mathbf{x}_\mu, \nu_\mu) \int \D\kappa\, \omega_\kappa \Phi^{(j_\mu)}_\kappa(\mathbf{x}_\mu) q_\kappa, \label{def:B2}\\
[A\mathbf{X}](\mu) =& \tilde{\alpha}(\mathbf{x}_\mu, \nu_\mu) \int \D\mu' \tilde{\alpha}(\mathbf{x}_{\mu}, \nu_{\mu'}) \notag \\
&\times \delta_{j_\mu j_{\mu'}}\delta(\mathbf{x}_\mu - \mathbf{x}_{\mu'}) X^{(j_{\mu'})}(\mathbf{x}_{\mu'}, \nu_{\mu'})\notag \\
= &\tilde{\alpha}(\mathbf{x}_\mu, \nu_\mu) \int \D\nu_{\mu'} \tilde{\alpha}(\mathbf{x}_{\mu}, \nu_{\mu'})X^{(j_\mu)}(\mathbf{x}_\mu, \nu_{\mu'}) .\label{def:A}
\end{align}
\end{subequations}
Here, we introduce the multi-index $\mu \equiv (\mathbf{x}_\mu, \nu_\mu, j_\mu)$ for the medium, where $\nu$ is the frequency of the excitation at position $\mathbf{x}_\mu$ inside the medium, and $j_\mu \in \{1, 2, 3\}$ labels the three components of the medium variables. The notation of the integration over $\mu$ is
\begin{equation}
    \int \D\mu := \int_0^\infty \D\nu_\mu \int_V \D^3x_\mu \sum_{j_\mu=1}^3,
\end{equation}
and the delta-function means $\delta(\mu - \mu') = \delta(\mathbf{x}_\mu - \mathbf{x}_{\mu'})\delta(\nu_\mu - \nu_{\mu'})\delta_{j_\mu j_{\mu'}}$.

The full Hamiltonian reads
\begin{equation}
    H = \frac{1}{2}
    \left(\begin{matrix}
        p \\
        \mathbf{\Pi}_X \\
        q \\
        \mathbf{X}
    \end{matrix}\right|
    \begin{matrix}
        \mathbb{I}& 0 & 0 & 0 \\
        0 & \mathbb{I}& 0 & 0 \\
        0 & 0 & \mathbb{I}_{\omega^2_\kappa} & B_1 \\
        0 & 0 & B_2 & \mathbb{I}_{\nu^2_\kappa} + A
    \end{matrix}
    \left|\begin{matrix}
        p \\
        \mathbf{\Pi}_X \\
        q \\
        \mathbf{X}
    \end{matrix}\right).
\end{equation}
Here, the identity integral operator is
\begin{equation}
    [\mathbb{I}f](\eta) = \int \D \eta'\,\delta(\eta-\eta')f(\eta')=f(\eta),
\end{equation}
and the frequency integral operators are
\begin{align}
    [\mathbb{I}_{\omega^2}f_{\kappa'}](\kappa) &= \int \D \kappa'\, \omega^2_{\kappa'}\delta(\kappa-\kappa')f_{\kappa'}=\omega^2_{\kappa}f_\kappa, \\
    [\mathbb{I}_{\nu^2}f_{\mu'}](\mu) &= \int \D \mu'\, \nu^2_{\mu'}\delta(\mu-\mu')f_{\mu'}=\nu^2_{\mu}f_\mu.
\end{align}
Introducing the variables $\underline{\mathbf{P}}$ and $\underline{\mathbf{Q}}$, constructed as the direct sum of the canonical variables $p,\, \mathbf{\Pi}_{X}$ and $q,\,\mathbf{X}$ respectively
\begin{equation}
\underline{\mathbf{P}}  = \begin{pmatrix}
p\\
\mathbf{\Pi}_X
\end{pmatrix}, \quad
\underline{\mathbf{Q}} = \begin{pmatrix}
q \\
\mathbf{X}
\end{pmatrix},
\end{equation}
we rewrite the Hamiltonian in the harmonic oscillator-like form as follows
\begin{equation}\label{eqn:Ham_matrix}
    H = \frac{1}{2}\left(\begin{matrix}
        \underline{\mathbf{P}} \\
        \underline{\mathbf{Q}}
    \end{matrix}\right|
    \begin{matrix}
        \mathbb{I}& 0 \\
        0 & \Omega^2
    \end{matrix}
    \left|\begin{matrix}
        \underline{\mathbf{P}} \\
        \underline{\mathbf{Q}}
    \end{matrix}\right) = \frac{1}{2}(\underline{\mathbf{P}}|\underline{\mathbf{P}}) + \frac{1}{2}(\underline{\mathbf{Q}}|\Omega^2\underline{\mathbf{Q}})
\end{equation}
with the squared plasmonic-polaritonic frequency operator
\begin{equation}
    \Omega^2 := \begin{bmatrix}
\mathbb{I}_{\omega_\kappa^2} & B_1  \\
B_2& \mathbb{I}_{\nu_\mu^2} + A
\end{bmatrix}= \Omega^2_0 + V
\end{equation}
decomposed into the free operator $\Omega^2_0$ and the coupling operator $V$
\begin{equation}
\Omega^2_0 = \begin{bmatrix}
\mathbb{I}_{\omega_\kappa^2} & 0 \\
0& \mathbb{I}_{\nu_\mu^2}
\end{bmatrix}, \quad V = \begin{bmatrix}
0 & B_1  \\
B_2&  A
\end{bmatrix}.
\end{equation}

\subsection{Spectral structure of plasmonic squared frequency operator}
The diagonalization of the Hamiltonian (\ref{eqn:Ham_matrix}) is reduced to the diagonalization of $\Omega^2$. The spectrum of $\Omega^2$ is composed of two continua $\omega_\kappa^2 \in [0, \infty)$ and $\nu_\mu^2 \in [0, \infty)$. The corresponding continuum eigenfunctions are labeled with multi-index $\kappa = (\mathbf{k}, \sigma_\kappa, \zeta_\kappa)$ for the electromagnetic part of the spectrum and $\mu = (\mathbf{x}_\mu, \nu_\mu, j_\mu)$ for the medium part:
\begin{subequations}
    \begin{align}
        \Omega^2 \underline{\psi}^e_{\kappa} &= \omega^2_\kappa \underline{\psi}^e_{\kappa}, \\
        \Omega^2 \underline{\psi}^m_{\mu} &= \nu^2_\mu \underline{\psi}^m_{\mu}.
    \end{align}
\end{subequations}
The eigenfunctions have the same structure as the variables $\underline{\mathbf{P}}$ and $\underline{\mathbf{Q}}$,
\begin{equation}\label{def:eigenfunc}
    \underline{\psi}^e_{\kappa}(\kappa', \mu') = \begin{bmatrix}
        u^e_{\kappa}(\kappa') \\
        v^e_{\kappa}(\mu')
    \end{bmatrix}, \quad 
    \underline{\psi}^m_{\mu}  (\kappa', \mu')  = \begin{bmatrix}
        u^m_{\mu}(\kappa') \\ 
        v^m_{\mu}(\mu')
    \end{bmatrix},
\end{equation}
where the upper components $u^e_\kappa(\kappa')$ and $u^m_\mu(\kappa')$ of the eigenfunctions are associated with the electromagnetic field, and the lower ones $v^e_{\kappa}(\mu')$ and $v^m_{\mu}(\mu')$ with the medium in the uncoupling limit. 

\begin{figure}[t]
   \centering
   \includegraphics[width=0.5\textwidth]{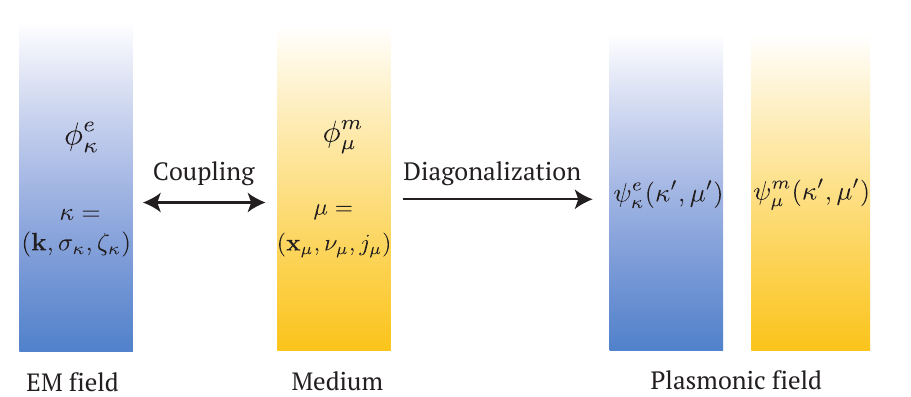} 
   \caption{The spectrum structure of the squared frequency plasmonic operators $\Omega^2 = \Omega^2_0 + V$ with the eigenfunctions of the uncoupled operator $\phi^e_\kappa$ and $\phi^m_\mu$ and the eigenfunctions of the coupled operator $\psi^e_\kappa$ and $\psi^m_\mu$. The unitarity of the diagonalization transformation preserves the double-spectrum and the degeneracy structure.}
   \label{fig:Diagonalization}
\end{figure}

The eigenfunctions are determined from the Lippmann-Schwinger equations \cite{Dorier2019}
\begin{subequations}\label{eq:LS_eq}
    \begin{align}
        \underline{\psi}^e_{\kappa} = \underline{\phi}^e_{\kappa} - \frac{1}{\Omega^2_0 - \omega^2_{\kappa} \mp i0^+} V \underline{\psi}^e_{\kappa}, \label{eq:LS_field}\\
        \underline{\psi}^m_{\mu} = \underline{\phi}^m_{\mu} - \frac{1}{\Omega^2_0 - \nu^2_{\mu} \mp i0^+} V \underline{\psi}^m_{\mu}, \label{eq:LS_medium}
    \end{align}
\end{subequations}
where $\underline{\phi}^e_{\kappa}$ and $\underline{\phi}^m_{\mu}$ are the eigenfunctions of the uncoupled operator $\Omega_0^2$, and $\mp i0^+$ describes the ways of avoiding the poles in the complex plane. Without loss of generality, from now on, we choose $-i0^+$. The eigenfunctions of the uncoupled operator $\Omega^2_0$ have the form 
\begin{equation}
\underline{\phi}^e_{\kappa}(\kappa')  = \begin{bmatrix}\delta(\kappa - \kappa') \\
0
\end{bmatrix} 
\end{equation}
for the electromagnetic field and
\begin{equation}
\underline{\phi}^m_{\mu}(\mu') = 
\begin{bmatrix}
0 \\
\delta(\mu - \mu')
\end{bmatrix}
\end{equation}
for the oscillators in the medium. The Lippmann-Schwinger equations provide a one-to-one unitary correspondence between the uncoupled and coupled modes. It was shown in \cite{Dorier2019}, \cite[Chapter 5]{lampart:tel-04769464}, that $\Omega_0^2$ and $\Omega^2$ are unitarily equivalent. Thus, their spectra have the same structure. To summarize, we depict the diagonalization scheme of the double spectrum structure in Figure \ref{fig:Diagonalization}.

\subsection{Complex representation of the classical Hamiltonian system}
The classical Hamiltonian can be rewritten as
\begin{equation}
\label{HamPsi}
    H = \frac{\hbar}{2} \left( \begin{matrix} 
    \mathbf{\Psi} \\
    \mathbf{\Psi}^{\ast}
   \end{matrix} \right| 
    \begin{matrix}
    \Omega & 0&  \\
    0& \Omega \\
    \end{matrix} \left| \begin{matrix}
    \mathbf{\Psi}^{\ast} \\
    \mathbf{\Psi}
    \end{matrix}\right)
\end{equation}
in terms of the complex field configurations $\underline{\mathbf{\Psi}}$ 
\begin{equation}
\label{PsiDef}
\underline{\mathbf{\Psi}}=\frac{1}{\sqrt{2\hbar}}\bigl(\Omega^{1/2}\underline{\mathbf{Q}}+i \Omega^{-1/2}\underline{\mathbf{P}}\bigr),
\end{equation}
defined in the following Hilbert space
\begin{align}
\mathcal{H} := \Bigg\{&\underline{\mathbf{\Psi}} = 
\begin{bmatrix}
u(\mathbf{k}, \sigma, \zeta) \\
v(\mathbf{x}, \nu, j)
\end{bmatrix}, \mathbf{k} \in \mathbb{R}^2\times[0,\infty), \sigma \in \{-, +\}, \notag \\
& \zeta \in \{c, s\}, \mathbf{x} \in V \subset \mathbb{R}^3, \nu \in (0, \infty), j = 1,2,3, \notag \\
& \text{ with } ( \underline{\mathbf{\Psi}} | \underline{\mathbf{\Psi}} ) < \infty \Bigg\}.
\end{align}
The classical Maxwell equations in this complex representation take the form
\begin{equation}
i\frac{\partial}{\partial t}\underline{\mathbf{\Psi}}=\Omega \underline{\mathbf{\Psi}},
\end{equation}
which has the same structure as a Schr\"odinger equation. We remark that we have incorporated $\hbar$ in the above definitions, despite their classical nature, for dimension reasons and to prepare the quantization.

\subsection{Diagonalization of the frequency operator $\Omega^2$}
Using the eigenfunctions (\ref{def:eigenfunc}), we can obtain the diagonalized frequency operator
\begin{equation}
\Omega_D = U^{\dagger} \Omega U= \left[ \begin{matrix} 
   \mathbb{I}_{\omega_{\kappa}} & 0\\
   0 & \mathbb{I}_{\nu_{\mu}}
    \end{matrix}\right],
\end{equation}
where
\begin{equation}\label{def:U}
    U =  \left[ \begin{matrix} 
   u^e_{\kappa} & u^m_{\mu}\\
   v^e_{\kappa} & v^m_{\mu}
    \end{matrix}\right].
\end{equation}
One can insert the diagonalizing matrix into the Hamiltonian:
\begin{equation}
\label{HamPsitilde}
    H = \frac{\hbar}{2} \left( \begin{matrix} 
   \underline{\tilde{\mathbf{\Psi}}} \\
       \underline{\tilde{\mathbf{\Psi}}}^{\ast}
   \end{matrix} \right| 
    \begin{matrix}
    \Omega_D & 0&  \\
    0& \Omega_D \\
    \end{matrix} \left| \begin{matrix}
       \tilde{\underline{\mathbf{\Psi}}} \\
       \tilde{\underline{\mathbf{\Psi}}}^{\ast}
    \end{matrix}\right),\qquad    \underline{\tilde{\mathbf{\Psi}}}=U^{\dagger}    \underline{ \mathbf{\Psi}},
\end{equation}
i.e.
  \begin{equation}
\label{HamPsitilde_}
    H = \frac{\hbar}{2} \!\int \!\!\D\kappa\,\omega_{\kappa} \bigl[ (\tilde z^e_{\kappa})^{\ast} \tilde z^e_{\kappa} + \mathrm{c.c.} \bigr] +  \frac{\hbar}{2} \!\int\!\!\D\mu\,\nu_{\mu} \bigl[ (\tilde z^m_{\mu})^{\ast} \tilde z^m_{\mu} + \mathrm{c.c.} \bigr],
\end{equation}
where we have denoted the components of $   \tilde{\underline{\mathbf{\Psi}}}$ as $
   \tilde{\underline{\mathbf{\Psi}} }= \begin{pmatrix}
\tilde z^e_\kappa\, \\ 
\tilde z^m_\mu 
\end{pmatrix}.
$

\subsection{Quantization of the Hamiltonian}
From the canonical quantization correspondence principle
  \begin{equation}\label{quantization}
  \tilde z^e_{\kappa}\rightarrow \hat C^e _{\kappa},\qquad   \tilde z^m_{\mu}\rightarrow \hat C^m _{\mu},
\end{equation}
we deduce the quantized Hamiltonian:
\begin{equation}
\label{HamCem}
\hat{H}=\int \D \kappa\, \hbar \omega_\kappa \hat{C}_\kappa^{e\dagger} \hat{C}^e_\kappa+ \int \D \mu\, \hbar \nu_\mu \hat{C}_\mu^{m\dagger} \hat{C}^m_\mu.
\end{equation}
It has the structure of two infinite harmonic oscillators associated in the uncoupling limit to the free field and the medium, respectively. They satisfy the canonical commutation relations
\begin{subequations}
\begin{align}
&\!\!\!\bigl[ \hat{C}^e_\kappa, \hat{C}^{e\dagger}_{\kappa'} \bigr] = \delta(\kappa - \kappa')\hat{\mathbb{I}}, \, \bigl[ \hat{C}^m_\mu, \hat{C}^{m\dagger}_{\mu'} \bigr] = \delta(\mu - \mu')\hat{\mathbb{I}}, \\
&\!\!\!\bigl[ \hat{C}^e_\kappa, \hat{C}^e_{\kappa'} \bigr] =\bigl[ \hat{C}^m_\mu, \hat{C}^m_{\mu'} \bigr] =\bigl[\hat{C}^e_\kappa, \hat{C}^m_\mu\bigr]=0.
\end{align}
\end{subequations}
The creation-annihilation operators are defined in the bosonic Fock space \cite{reed1972methods, debievre2006, debievre2007s, honegger2015photons}
\begin{equation}
\mathcal{F}^B(\mathcal{H}) = \bigoplus_{n=0}^\infty \mathcal{F}_n 
\end{equation}
with
\begin{subequations}
\begin{align}
\mathcal{F}_0^B &:= \mathbb{C}, \\
\mathcal{F}_0^B &:= \mathcal{H}, \\
\mathcal{F}_0^B &:= \mathcal{H} \otimes_s \mathcal{H}, \\
&\vdots \notag \\
\mathcal{F}_0^B &:= \mathcal{H} \otimes_s \mathcal{H} \dots  \otimes_s \mathcal{H}, \\
&\vdots \notag
\end{align}
\end{subequations}
The action of the creation operators on the vacuum $|\varnothing \rangle$ is 
\begin{subequations}
\begin{align}
\hat{C}^e_\kappa|\varnothing \rangle & = |\underline{\psi}^e_{\kappa} \rangle,\\
 \hat{C}^m_\mu |\varnothing \rangle & = |\underline{\psi}^m_{\mu} \rangle.
\end{align}
\end{subequations}
Thus, the single-quantum subspace is identified with the classical Hilbert space $\mathcal{H}$ of the system composed of the electromagnetic field and the medium, represented as an infinite collection of charged harmonic oscillators, which is meant as a simple linear model for the electrons and the phonons of the medium.

\section{Electric field operator}\label{sec:EM_field}
\subsection{Canonical observables operators in terms of the creation-annihilation operators}
Inverting the Eq.~(\ref{PsiDef}) and substituting the diagonalizing matrix $U$, we obtain the expression for the canonical variables
\begin{align}
    \underline{\mathbf{Q}} &= \sqrt{\frac{\hbar}{2}}(U\Omega^{-1/2}_D \tilde{\underline{\mathbf{\Psi}}} + U^*\Omega^{-1/2}_D \tilde{\mathbf{\Psi}}^*), \label{par:Q}\\
    \underline{\mathbf{P}} &= -i\sqrt{\frac{\hbar}{2}}(U\Omega^{1/2}_D \tilde{\underline{\mathbf{\Psi}}} - U^*\Omega^{1/2}_D \tilde{\mathbf{\Psi}}^*). \label{par:P}
\end{align}
Substituting the diagonalizing matrix $U$ from Eq.~(\ref{def:U}) and applying the correspondence principle as in Eq.~(\ref{quantization}), we obtain the canonical operators in terms of the creation-annihilation operators  
\begin{align}\label{QPhat}
\underline{\hat{\mathbf{Q}}} = \begin{bmatrix}
    \hat{q}\\
    \hat{\mathbf{X}}
\end{bmatrix} &= \int \D\kappa \sqrt{\frac{\hbar}{2\omega_{\kappa}}} \underline{\mathbf{\psi}}^e_{\kappa}(\kappa_q, \mu_X)\hat{C}^e_{\kappa} + \notag\\
&+ \int \D\mu \sqrt{\frac{\hbar}{2\nu_{\mu}}}\underline{\mathbf{\psi}}^m_{\mu}(\kappa_q, \mu_X)\hat{C}^m_{\mu} + \text{H.c.}\\
\underline{\hat{\mathbf{P}}} = \begin{bmatrix}
    \hat{p}\\
    \hat{\mathbf{\Pi}}_X
\end{bmatrix} &= -i\left[\int \D\kappa \sqrt{\frac{\hbar\omega_{\kappa}}{2}} \underline{\mathbf{\psi}}^e_{\kappa}(\kappa_p, \mu_P)\hat{C}^e_{\kappa}\right. \notag \\
&+ \left.\int \D \mu \sqrt{\frac{\hbar\nu_{\mu}}{2}} \underline{\mathbf{\psi}}^m_{\mu}(\kappa_p, \mu_P)\hat{C}^m_{\mu}\right] + \text{H.c.},
\end{align}
where $\underline{\psi}^e_{\kappa}(\kappa_q, \mu_X)$ and $\underline{\psi}^m_{\mu}(\kappa_q, \mu_X)$ have the structure given in Eq.~(\ref{def:eigenfunc}). 
One can obtain the commutation relations of the canonical operators using the commutation relations of the creation-annihilation operators and the orthonormality property of the eigenbasis of $\Omega_0^2$:
\begin{subequations}
\begin{align}
[\hat{q}, \hat{p}] &= i\hbar\delta(\kappa_q - \kappa_p),\\
[\hat{\mathbf{X}}, \hat{\mathbf{\Pi}}] &= i\hbar \delta(\mu_X - \mu_{\Pi}),\\
[\hat{q}, \hat{\mathbf{X}}] &= [\hat{p}, \hat{\mathbf{\Pi}}^\dagger] = 0.
\end{align}
\end{subequations}

\subsection{Electric field operator in terms of the coupled electromagnetic and medium eigenfunctions}
The quantized version of the electromagnetic field can be expressed using the classical definition from Eq.~(\ref{def:momentum_A})
\begin{equation}
\mathbf{E}(\mathbf{r}) = -\frac{1}{\varepsilon_0}\mathbf{\Pi}_A(\mathbf{r}) - \frac{1}{\varepsilon_0}\int_0^\infty\D\nu\, \alpha(\mathbf{r}, \nu)\mathbf{X}(\mathbf{r}, \nu). 
\end{equation}
Expressing the classical momentum of the field $\mathbf{\Pi}_A$ through the variables $q_\kappa$ as in Eq.~(\ref{par:momentum_A}) and replacing $q_\kappa \rightarrow \hat{q}_\kappa$, $\mathbf{X}_\mu \rightarrow \hat{\mathbf{X}}_\mu$, we obtain
\begin{align}
    \hat{\mathbf{E}}(\mathbf{r}) = -\frac{1}{\sqrt{\varepsilon_0}}&\left[ \int\D \kappa\, \omega_\kappa \mathbf{\Phi}_\kappa(\mathbf{r}) \hat{q}_\kappa\right. \notag \\
    &+ \left.\int_0^\infty \D \nu\, \tilde\alpha(\mathbf{r}, \nu) \hat{\mathbf{X}}(\mathbf{r}, \nu)\right].
\end{align}
Substituting the operators $\hat{q}_\kappa$  and $\hat{\mathbf{X}}_\mu$ from Eq.~(\ref{QPhat}), we express the field operator in terms of the creation-annihilation operators
\begin{equation}
\hat{\mathbf{E}}(\mathbf{r}) = \hat{\mathbf{E}}^e(\mathbf{r}) + \hat{\mathbf{E}}^m(\mathbf{r}),
\end{equation}
where
\begin{align}
\hat{\mathbf{E}}^e(\mathbf{r}) &= -\int \D\kappa \sqrt{\frac{\hbar}{2\varepsilon_0 \omega_\kappa }}\left[ \mathbf{e}_\kappa(\mathbf{r}) \hat{C}^e_\kappa + \text{H.c.}\right], \label{def:Ee} \\
 \hat{\mathbf{E}}^m(\mathbf{r}) &= -\int \D\mu \sqrt{\frac{\hbar}{2\varepsilon_0 \nu }}\left[ \mathbf{m}_\mu(\mathbf{r}) \hat{C}^m_\mu + \text{H.c.}\right]. \label{def:Em}
\end{align}
We call $\mathbf{e}_\kappa(\mathbf{r})$ and $\mathbf{m}_\mu(\mathbf{r})$ the field coefficients. They are defined through the eigenfunctions $\underline{\psi}^e_\kappa = \begin{pmatrix} u^e_{\kappa}(\kappa') \\ v^e_{\kappa}(\mu') \end{pmatrix}$ and $\underline{\psi}^m_\mu = \begin{pmatrix} u^m_{\mu}(\kappa') \\ v^m_{\mu}(\mu') \end{pmatrix}$ of $\Omega^2$ as follows
\begin{align}
\mathbf{e}_\kappa(\mathbf{r}) &= \mathbf{e}^u_\kappa(\mathbf{r}) + \mathbf{e}^v_\kappa(\mathbf{r}), \label{def:e} \\
\mathbf{m}_\kappa(\mathbf{r}) &= \mathbf{m}^u_\mu(\mathbf{r}) + \mathbf{m}^v_\mu(\mathbf{r}), \label{def:m}
\end{align}
where
\begin{align}
\mathbf{e}^u_\kappa(\mathbf{r}) &= \int \D\kappa' \omega_{\kappa'} \mathbf{\Phi}_{\kappa'}(\mathbf{r})u^e_{\kappa}(\kappa'), \label{def:eu}\\
\mathbf{e}^v_\kappa(\mathbf{r}) &=  \int_0^\infty \D\nu' \tilde{\alpha}(\mathbf{r}, \nu')\mathbf{v}^e_{\kappa}(\mathbf{r}, \nu'), \label{def:ev}\\
\mathbf{m}^u_\mu(\mathbf{r}) &=  \int \D\kappa' \omega_{\kappa'} \mathbf{\Phi}_{\kappa'}(\mathbf{r})u^m_{\mu}(\kappa'), \label{def:mu}\\
\mathbf{m}^v_\mu(\mathbf{r}) &=  \int_0^\infty \D\nu' \tilde{\alpha}(\mathbf{r}, \nu')\mathbf{v}^m_{\mu}(\mathbf{r}, \nu'). \label{def:mv}
\end{align}
We remark that $\mathbf{e}^u_\kappa(\mathbf{r})$ and $\mathbf{m}^u_\mu(\mathbf{r})$ are transverse fields, while $\mathbf{e}^v(\mathbf{r})$ and $\mathbf{m}^v(\mathbf{r})$ have a transverse and a longitudinal components.

The eigenfunctions $u^e_{\kappa}(\kappa'),\, v^e_{\kappa}(\mu'),\, u^m_{\mu}(\kappa'),\, v^m_{\mu}(\mu')$ are determined by the Lippmann-Schwinger equations (\ref{eq:LS_eq}) with $v^e_{\kappa}(\mu') = v^e_{\kappa}(\mathbf{x}', \nu', j')$ and $v^m_{\mu}(\mu') = v^m_{\mu}(\mathbf{x}', \nu', j')$. We use the following vectorial notation
\begin{subequations}
\begin{align}
    \mathbf{v}^e_{\kappa}(\mathbf{x}', \nu') &:= \sum_{j'}v^e_{\kappa}(\mathbf{x}', \nu', j')\mathbf{n}_{j'}, \label{def:v_e_vec} \\
    \mathbf{v}^m_{\mu}(\mathbf{x}', \nu') &:= \sum_{j'}v^m_{\mu}(\mathbf{x}', \nu', j')\mathbf{n}_{j'},\label{def:v_m_vec}
\end{align}    
\end{subequations}
where $\mathbf{n}_{j'}$ is the unit vector of the corresponding coordinate $j'$. The explicit form of the Lippmann-Schwinger equations in terms of the upper and lower components of the eigenfunctions is
\begin{subequations}\label{eqn:LS_e}
\begin{align}
u^e_{\kappa}(\kappa') &= \delta(\kappa - \kappa') - \frac{\left[ B_1v^e_{\kappa} \right](\kappa')}{\omega_{\kappa'}^2 - \omega_{\kappa}^2 - i0} \label{eqn:LS_u_e}\\
v^e_{\kappa}(\mu') &= - \frac{\left[ B_2u^e_{\kappa} \right](\mu') + \left[ A v^e_{\kappa} \right](\mu')}{\nu_{\mu'}^2 - \omega_{\kappa}^2 - i0}, \label{eqn:LS_v_e}
\end{align}
\end{subequations}
for the eigenfunctions $\underline{\psi}^e_\kappa = \begin{pmatrix} u^e_{\kappa}(\kappa') \\ v^e_{\kappa}(\mu') \end{pmatrix}$ of $\Omega^2$ related to electromagnetic field and
\begin{subequations}\label{eqn:LS_m}
\begin{align}
u^m_{\mu}(\kappa') &=  - \frac{\left[ B_1v^m_{\mu} \right](\kappa')}{\omega_{\kappa'}^2 - \nu_{\mu}^2 - i0}\label{eqn:LS_u_m} \\
v^m_{\mu}(\mu') &= \delta(\mu - \mu') - \frac{\left[ B_2u^m_{\mu} \right](\mu') + \left[ Av^m_{\mu} \right](\mu')}{\nu_{\mu'}^2 - \nu_{\mu}^2 - i0}
\label{eqn:LS_v_m}
\end{align}
\end{subequations}
for the eigenfunctions $\underline{\psi}^m_\mu = \begin{pmatrix} u^m_{\mu}(\kappa') \\ v^m_{\mu}(\mu') \end{pmatrix}$ of $\Omega^2$ related to the medium in the uncoupling limit.

\subsection{Integral equation for the $\mathbf{e}$ coefficients}

We will show, that the $\mathbf{e}$ coefficient, introduced in (\ref{def:Ee}), satisfies the following integral equation
\begin{align}\label{eqn:int_e}
\mathbf{e}_\kappa(\mathbf{r}) =&  \omega_{\kappa}\mathbf{\Phi}_\kappa(\mathbf{r}) \notag \\
+& \frac{\omega^2_\kappa}{c^2} \int_V \D^3z\,\bar{\bar{G}}_{0+}(\mathbf{r}, \mathbf{z}, \omega_\kappa) [\varepsilon(\mathbf{z}, \omega_\kappa)  - 1]\mathbf{e}_\kappa(\mathbf{z})
\end{align}
with the dyadic Green tensor of the free field $\bar{\bar{G}}_{0+}(\mathbf{r}, \mathbf{z}, \omega_\kappa)$, that satisfies
\begin{equation}\label{eqn:Green_tensor}
\nabla \times \nabla \times \bar{\bar{G}}_{0+}(\mathbf{r}, \mathbf{z}, \omega_\kappa) - \frac{\omega^2_\kappa}{c^2}\bar{\bar{G}}_{0+}(\mathbf{r}, \mathbf{z}, \omega_\kappa) = \bar{\bar{I}}\delta(\mathbf{r} - \mathbf{z}),
\end{equation} 
and the Sommerfeld outgoing radiation condition \cite{sommerfeld1949partial} 
\begin{equation}\label{eqn:Sommerfeld_cond}
\lim_{|\mathbf{r}| \rightarrow \infty }|\mathbf{r}|\left( \nabla \times - \frac{i\omega_\kappa}{c} \frac{\mathbf{r}}{|\mathbf{r}|} \times \right)\bar{\bar{G}}_{0+}(\mathbf{r}, \mathbf{z}, \omega_\kappa) = 0.
\end{equation}
One can obtain Eq.~(\ref{eqn:int_e}) from the Lippmann-Schwinger equation (\ref{eqn:LS_e}) as follows.

We first consider Eq.~(\ref{eqn:LS_u_e}). Multiplying it by $\D\kappa'\, \omega_{\kappa'} \mathbf{\Phi}_{\kappa'}(\mathbf{r})$ and then integrating it, we obtain
\begin{align}
\mathbf{e}^u_{\kappa}(\mathbf{r}) =& \omega_{\kappa}\mathbf{\Phi}_\kappa(\mathbf{r})\notag \\
&- \int \D\kappa' \frac{\omega_{\kappa'}\mathbf{\Phi}_{\kappa'}(\mathbf{r})}{\omega^2_{\kappa'} - \omega^2_\kappa - i0^+}\left[ B_1v^e_{\kappa} \right](\kappa'). \label{eqn:e_first}
\end{align}
Using the definition of the integral operator $\left[ B_1v^e_{\kappa} \right](\kappa')$ (\ref{def:B1}) and the definition of $\mathbf{e}^v_\kappa(\mathbf{r})$ (\ref{def:ev}), we rewrite the integral term as follows
\begin{align}\label{eqn:int_b1_e}
    & \int \D\kappa' \frac{\omega_{\kappa'}\mathbf{\Phi}_{\kappa'}(\mathbf{r})}{\omega^2_{\kappa'} - \omega^2_\kappa - i0^+}\left[ B_1v^e_{\kappa} \right](\kappa') = \notag \\ 
    &=  \int_{\mathbb{R}^3} \D^3z\, \int \D \kappa' \, \frac{\omega^2_{\kappa'}\mathbf{\Phi}_{\kappa'}(\mathbf{r}) \otimes \mathbf{\Phi}_{\kappa'}(\mathbf{z})}{\omega^2_{\kappa'} - \omega^2_\kappa - i0^+}  \mathcal{P}^\perp\mathbf{e}^v_\kappa(\mathbf{z}).
\end{align}
One can express the factor$
\int \D\kappa' \, \frac{\omega^2_{\kappa'}\mathbf{\Phi}_{\kappa'}(\mathbf{r}) \otimes \mathbf{\Phi}_{\kappa'}(\mathbf{z})}{\omega^2_{\kappa'} - \omega^2_\kappa - i0^+} $ through the transverse Green tensor
\begin{align}
&\int \D\kappa' \, \frac{\omega^2_{\kappa'}\mathbf{\Phi}_{\kappa'}(\mathbf{r}) \otimes \mathbf{\Phi}_{\kappa'}(\mathbf{z})}{\omega^2_{\kappa'} - \omega^2_\kappa - i0^+} \notag \\
&= \int \D\kappa' \, \frac{(\omega^2_{\kappa'} - \omega^2_\kappa + \omega^2_\kappa)\mathbf{\Phi}_{\kappa'}(\mathbf{r}) \otimes \mathbf{\Phi}_{\kappa'}(\mathbf{x})}{\omega^2_{\kappa'} - \omega^2_\kappa - i0^+} \notag \\
&= \int\D\kappa' \,\left[\omega^2_\kappa\frac{\mathbf{\Phi}_{\kappa'}(\mathbf{r}) \otimes \mathbf{\Phi}_{\kappa'}(\mathbf{z})}{\omega^2_{\kappa'} - \omega^2_\kappa - i0^+} +  \mathbf{\Phi}_{\kappa'}(\mathbf{r}) \otimes \mathbf{\Phi}_{\kappa'}(\mathbf{z})\right] \notag \\
&= \frac{\omega^2_\kappa}{c^2}\bar{\bar{G}}^\perp_{0+}(\mathbf{r}, \mathbf{z}, \omega_\kappa) + \delta^\perp(\mathbf{r} - \mathbf{z}), 
\end{align}
where $\bar{\bar{G}}^\perp_{0+}(\mathbf{r}, \mathbf{z}, \omega_\kappa)$ is defined as
\begin{equation}
\bar{\bar{G}}^\perp_{0+}(\mathbf{r}, \mathbf{z}, \omega_\kappa) := \int\D\kappa' \,c^2 \frac{\mathbf{\Phi}_{\kappa'}(\mathbf{r}) \otimes \mathbf{\Phi}_{\kappa'}(\mathbf{z})}{\omega^2_{\kappa'} - \omega^2_\kappa - i0^+}.
\end{equation}
Here, we used the completeness of the basis to obtain $\delta^\perp(\mathbf{r} - \mathbf{x})$ and the definition of the Green tensor of the vector Laplacian with the transversality constraint. We discuss the relation between the transverse  Green tensor of the free field with the complete one $\bar{\bar{G}}_{0+}(\mathbf{r}, \mathbf{z}, \omega_\kappa)$ in Appendix \ref{app:Green}. The final expression of the integral (\ref{eqn:int_b1_e}) reads
\begin{align}\label{transverse_e}
&\int \D\kappa' \frac{\omega_{\kappa'}\mathbf{\Phi}_{\kappa'}(\mathbf{r})}{\omega^2_{\kappa'} - \omega^2_\kappa - i0^+}\left[ B_1v^e_{\kappa} \right](\kappa') \notag \\
&= \int_{\mathbb{R}^3} \D^3z\,\left[ \frac{\omega^2_\kappa}{c^2}\bar{\bar{G}}^\perp_{0+}(\mathbf{r}, \mathbf{z}, \omega_\kappa)+ \delta^\perp(\mathbf{r} - \mathbf{z})\right]\mathbf{e}^v_\kappa(\mathbf{z})\notag \\
&= \mathcal{P}^\perp\mathbf{e}^v_\kappa(\mathbf{r}) +  \frac{\omega^2_\kappa}{c^2}\int_{\mathbb{R}^3} \D^3z\,\bar{\bar{G}}^\perp_{0+}(\mathbf{r}, \mathbf{z}, \omega_\kappa)\mathbf{e}^v_\kappa(\mathbf{z}).
\end{align}
Using (\ref{transverse_e}), we rewrite (\ref{eqn:e_first}) as
\begin{align}
\mathbf{e}^u_{\kappa}(\mathbf{r}) &= \omega_{\kappa}\mathbf{\Phi}_\kappa(\mathbf{r}) - \mathcal{P}^\perp\mathbf{e}^v_{\kappa}(\mathbf{r}) \notag \\
 &-\frac{\omega^2_\kappa}{c^2}  \int_{\mathbb{R}^3} \D^3z\,\bar{\bar{G}}^\perp_{0+}(\mathbf{r}, \mathbf{z}, \omega_\kappa)\mathbf{e}^v_{\kappa}(\mathbf{z}).\label{eqn:euk}
\end{align}

We now consider Eq.~(\ref{eqn:LS_v_e}). One can multiply it by $\D\nu'\, \tilde{\alpha}(\mathbf{r}, \nu')\mathbf{n}_{j'}$, integrate it over all frequencies, sum over $j'$, and use the definition (\ref{def:ev}). It yields
\begin{align}
\mathbf{e}^v_{\kappa}(\mathbf{r}) =& \int \D \nu'  \sum_{j'}\tilde{\alpha}(\mathbf{r}, \nu') v^e_\kappa(\mathbf{r}, \nu', j')\mathbf{n}_j\notag \\
=&- \int \D\nu' \sum_{j'}\frac{\tilde{\alpha}(\mathbf{r}, \nu')\left[ B_2u^e_{\kappa} \right](\mathbf{r}, \nu', j')}{\nu'^2 - \omega_{\kappa}^2 - i0}\mathbf{n}_{j'}\notag \\ 
&- \int \D\nu' \sum_{j'}\frac{\tilde{\alpha}(\mathbf{r}, \nu')\left[ A v^e_{\kappa} \right](\mathbf{r}, \nu', j')}{\nu'^2 - \omega_{\kappa}^2 - i0}\mathbf{n}_{j'}.\label{eqn:e_second}
\end{align}
To calculate the integral terms, we use the definitions of the integral operators $\left[ B_2u^e_{\kappa} \right](\mathbf{r}, \nu)$ and $\left[ A v^e_{\kappa} \right](\mathbf{r}, \nu)$, given in (\ref{def:B2}) and (\ref{def:A}), the Kramers-Kronig relation (\ref{eqn:KK}), and the definitions of $\mathbf{e}^u_\kappa(\mathbf{r})$ and $\mathbf{e}^v_\kappa(\mathbf{r})$, given in (\ref{def:eu}) and (\ref{def:ev}). They read
\begin{align}\label{eps_eu}
&\int \D\nu'\frac{\tilde{\alpha}(\mathbf{r}, \nu')}{\nu'^2 - \omega_{\kappa}^2 - i0}\sum_{j'}\left[ B_2u^e_{\kappa} \right](\mathbf{r}, \nu', j')\mathbf{n}_{j'} \notag \\
&= \int \D \nu'\frac{\tilde{\alpha}(\mathbf{r}, \nu')}{\nu'^2 - \omega_{\kappa}^2 - i0} \tilde{\alpha}(\mathbf{r}, \nu') \int \D \kappa'\, \omega_{\kappa'} \mathbf{\Phi}_{\kappa'}(\mathbf{r})u^e_{\kappa}(\kappa') \notag \\
&= \left[ \varepsilon(\mathbf{r}, \omega_\kappa) - 1\right]\mathbf{e}^u_\kappa(\mathbf{r}),
\end{align}
where $\mathbf{\Phi}_{\kappa'}(\mathbf{r}) = \sum_{j'}\Phi^{(j')}_{\kappa'}(\mathbf{r})\mathbf{n}_{j'}$, and
\begin{align}\label{eps_ev}
&\int \D \nu'\frac{\tilde{\alpha}(\mathbf{r}, \nu')}{\nu'^2 - \omega_{\kappa}^2 - i0}\sum_{j'}\left[ A v^e_{\kappa} \right](\mathbf{r}, \nu', j')\mathbf{n}_{j'} \notag \\
&= \int \D\nu'\frac{\tilde{\alpha}(\mathbf{r}, \nu')}{\nu'^2 - \omega_{\kappa}^2 - i0} \tilde{\alpha}(\mathbf{r}, \nu')\, \int_0^\infty \D\nu'\, \tilde{\alpha}(\mathbf{r}, \nu')\mathbf{v}^e_\kappa(\mathbf{r}, \nu') \notag \\
&= \left[ \varepsilon(\mathbf{r}, \omega_\kappa) - 1\right]\mathbf{e}^v_\kappa(\mathbf{r}),
\end{align}
where we used (\ref{def:v_e_vec}). Substitution of the two latter expressions into (\ref{eqn:e_second}) yields
\begin{align}
\mathbf{e}^v_{\kappa}(\mathbf{r}) &= -[\varepsilon(\mathbf{r}, \omega_\kappa) - 1]\left[\mathbf{e}^u_{\kappa}(\mathbf{r}) + \mathbf{e}^v_{\kappa}(\mathbf{r})\right] \notag \\
&=  -[\varepsilon(\mathbf{r}, \omega_\kappa) - 1]\mathbf{e}_{\kappa}(\mathbf{r}). \label{eqn:evk}
\end{align}

After obtaining the relations (\ref{eqn:euk}), (\ref{eqn:evk}) for $\mathbf{e}^u_{\kappa}(\mathbf{r})$ and $\mathbf{e}^v_{\kappa}(\mathbf{r})$, we can derive Eq.~(\ref{eqn:int_e}). Substituting (\ref{eqn:evk}) into the integral in (\ref{eqn:euk}) and transfering $\mathcal{P}^\perp\mathbf{e}^v_{\kappa}(\mathbf{r})$ to the left side of the equation, we have
\begin{align}
\mathbf{e}^u_{\kappa}(\mathbf{r}) &+ \mathcal{P}^\perp\mathbf{e}^v_{\kappa}(\mathbf{r}) = \omega_{\kappa}\mathbf{\Phi}_\kappa(\mathbf{r}) \notag \\
&+ \frac{\omega^2_\kappa}{c^2}  \int_V \D^3z\,\bar{\bar{G}}^\perp_{0+}(\mathbf{r}, \mathbf{z}, \omega_\kappa)[\varepsilon(\mathbf{z}, \omega_\kappa)  - 1]\mathbf{e}_{\kappa}(\mathbf{z}).\label{eqn:eu+ev_perp}
\end{align}
Also, from (\ref{eqn:evk}), we can obtain the equation for the longitudinal component of $\mathbf{e}^v_{\kappa}(\mathbf{r})$ as follows
\begin{align}
\mathcal{P}^\parallel\mathbf{e}^v_{\kappa}(\mathbf{r}) &= -\int_{\mathbb{R}^3} \D^3 x\, \delta^\parallel(\mathbf{r} - \mathbf{z})\left[\varepsilon(\mathbf{z}, \omega_\kappa)  - 1\right]\mathbf{e}_{\kappa}(\mathbf{z}) \notag \\
& =  \frac{\omega_\kappa^2}{c^2}\int_V  \D^3 z\, \bar{\bar{G}}^\parallel_{0+}(\mathbf{r}, \mathbf{z}, \omega_\kappa)\left[\varepsilon(\mathbf{z}, \omega_\kappa)  - 1\right]\mathbf{e}_{\kappa}(\mathbf{z}).\label{eqn:ev_parallel}
\end{align}
Here, $\mathcal{P}^\parallel$ is the projector on the longitudinal fields, and $\delta^\parallel(\mathbf{r} - \mathbf{z})$ is the longitudinal delta-function. We used the expression that links the longitudinal delta-function and the longitudinal component of the free Green tensor $\bar{\bar{G}}^\parallel_{0+}(\mathbf{r}, \mathbf{z}, \omega_\kappa)$ as (see Appendix \ref{app:Green})
\begin{equation}\label{eqn:delta_long}
\delta^\parallel(\mathbf{r} - \mathbf{z}) = -\frac{\omega_\kappa^2}{c^2}\bar{\bar{G}}^\parallel_{0+}(\mathbf{r}, \mathbf{z}, \omega_\kappa).
\end{equation}
Adding (\ref{eqn:eu+ev_perp}) and (\ref{eqn:ev_parallel}), we obtain an integral expression for $\mathbf{e}_\kappa(\mathbf{r})$
\begin{align} \label{integral-eq-e-hrj}
\mathbf{e}_\kappa(\mathbf{r}) =& \mathbf{e}^u_\kappa(\mathbf{r}) + \mathbf{e}^v_\kappa(\mathbf{r}) \notag \\
=& \mathbf{e}^u_\kappa(\mathbf{r}) + \mathcal{P}^\perp\mathbf{e}^v_{\kappa}(\mathbf{r}) + \mathcal{P}^\parallel\mathbf{e}^v_{\kappa}(\mathbf{r}) \notag \\
=& \omega_{\kappa}\mathbf{\Phi}_\kappa(\mathbf{r}) + \frac{\omega^2_\kappa}{c^2} \int_V \D^3z\,\bar{\bar{G}}_{0+}(\mathbf{r}, \mathbf{z}, \omega_\kappa)\notag \\
&\times[\varepsilon(\mathbf{z}, \omega_\kappa)  - 1] \mathbf{e}_\kappa(\mathbf{z}),
\end{align}
where we also used $\mathbf{e}^v_{\kappa}(\mathbf{r}) = \mathcal{P}^\perp\mathbf{e}^v_{\kappa}(\mathbf{r}) + \mathcal{P}^\parallel\mathbf{e}^v_{\kappa}(\mathbf{r})$, $\bar{\bar{G}}_{0+}(\mathbf{r}, \mathbf{z}, \omega_\kappa) = \bar{\bar{G}}^\perp_{0+}(\mathbf{r}, \mathbf{z}, \omega_\kappa) + \bar{\bar{G}}^\parallel_{0+}(\mathbf{r}, \mathbf{z}, \omega_\kappa)$ (see Appendix \ref{app:Green}) and (\ref{def:e}). Since the function $\varepsilon(\mathbf{z}, \omega_\kappa) - 1$ is zero outside $V$, we can restrict the volume of integration with $V$, which concludes the derivation of Eq.~(\ref{eqn:int_e}).

\subsection{Integral equation for the $\mathbf{m}$ coefficients}

We will show that the $\mathbf{m}$ coefficients, introduced in (\ref{def:Em}), satisfy the following integral equation
\begin{align}\label{eqn:m_int}
\mathbf{m}_{\mu}(\mathbf{r}) &=  - \tilde{\alpha}( \mathbf{r}, \nu_\mu)\frac{\nu^2}{c^2} \bar{\bar{G}}_{0+}(\mathbf{r}, \mathbf{x}_\mu, \nu_\mu)\mathbf{n}_{j_\mu} \notag \\
&+ \frac{\nu_\mu^2}{c^2}\int_V \D^3z\,\bar{\bar{G}}_{0+}(\mathbf{r}, \mathbf{z}, \nu) [\varepsilon(\mathbf{z}, \nu_\mu) - 1]\mathbf{m}_{\mu}(\mathbf{z}).
\end{align}
One can obtain this expression from the Lippmann-Schwinger equation (\ref{eqn:LS_m}) for the eigenfunctions $\underline{\psi}^m_\mu$ associated with the medium as follows.

First, we insert (\ref{eqn:LS_u_m}) into the definition (\ref{def:mu}) of $\mathbf{m}^u_\mu(\mathbf{r})$
\begin{align}
\mathbf{m}^u_{\mu}(\mathbf{r}) :=&\int \D\kappa\, \omega_\kappa \mathbf{\Phi}_\kappa(\mathbf{r}) u^m_\mu(\kappa) \notag \\
&- \int \D\kappa' \frac{\omega_{\kappa'}\mathbf{\Phi}_{\kappa'}(\mathbf{r})}{\omega^2_{\kappa'} - \nu^2_\mu - i0^+}\left[ B_1v^m_{\mu} \right](\kappa').
\end{align}
Inserting (\ref{eqn:LS_v_m}) into the definition (\ref{def:mv}) of $\mathbf{m}^v_\mu(\mathbf{r})$ yields
\begin{align}
\mathbf{m}^v_{\mu}(\mathbf{r}) =& \int \D\nu'\sum_{j'} \tilde{\alpha}(\mathbf{r}, \nu') v^m_\mu(\mathbf{r}, \nu', j') \notag \\
=& \int \D\nu'\sum_{j'} \tilde{\alpha}(\mathbf{r}, \nu')\delta(\mathbf{x}_\mu - \mathbf{r})\delta(\nu_\mu - \nu')\delta_{j_\mu j'}\mathbf{n}_{j'} \notag \\
&- \int \D\nu'\sum_{j'}\frac{\tilde{\alpha}(\mathbf{r}, \nu')\left[ B_2u^m_{\mu} \right](\mathbf{r}, \nu', j') }{\nu'^2 - \nu^2_\mu - i0}\mathbf{n}_{j'} \notag \\
&- \int \D\nu'\sum_{j'}\frac{\tilde{\alpha}(\mathbf{r}, \nu')\left[ A v^m_{\mu} \right](\mathbf{r}, \nu', j')}{\nu'^2 - \nu^2_\mu - i0}\mathbf{n}_{j'}.
\end{align}
The integral-operator terms are the same as in the equations for $\mathbf{e}^u_\kappa(\mathbf{r})$ and $\mathbf{e}^v_{\kappa}(\mathbf{r})$ replacing $u^e_\kappa$ and $v^e_\kappa$ by $u^m_\mu$ and $v^m_\mu$ in (\ref{eps_eu}) and (\ref{eps_ev}). The final forms of the equations are
\begin{align}
\mathbf{m}^u_{\mu}(\mathbf{r}) =&  - \mathcal{P}^\perp\mathbf{m}^v_{\mu}(\mathbf{x}) \notag \\
&- \frac{\nu^2_\mu}{c^2} \int_{\mathbb{R}^3} \D^3z\,\bar{\bar{G}}^\perp_{0+}(\mathbf{r}, \mathbf{z}, \nu_\mu)\mathbf{m}^v_{\mu}(\mathbf{z}), \label{eqn:mu}\\
\mathbf{m}^v_{\mu}(\mathbf{r}) =& \tilde{\alpha}( \mathbf{r}, \nu_\mu)\delta(\mathbf{r} - \mathbf{x}_\mu)\mathbf{n}_{j_\mu} - [\varepsilon( \mathbf{r}, \nu_\mu) - 1]\,\mathbf{m}_{\mu}(\mathbf{r}).\label{eqn:mv}
\end{align}
Substituting (\ref{eqn:mv}) into the integral in (\ref{eqn:mu}) and taking $\mathcal{P}^\perp\mathbf{m}^v_{\mu}(\mathbf{x})$ to the left side of the equation, we obtain
\begin{align}\label{eqn:mu+mv_perp}
\mathbf{m}^u_{\mu}(\mathbf{r}) &+ \mathcal{P}^\perp\mathbf{m}^v_{\mu}(\mathbf{r}) =  -\frac{\nu^2}{c^2}\bar{\bar{G}}^\perp_{0}(\mathbf{r}, \mathbf{x}_\mu, \nu_\mu) \tilde{\alpha}(\mathbf{x}_\mu, \nu_\mu)\mathbf{n}_{j_\mu} \notag \\
&+ \frac{\nu^2_\mu}{c^2}\int_V \D^3 z\, \bar{\bar{G}}^\perp_{0+}(\mathbf{r}, \mathbf{z}, \nu_\mu) [\varepsilon(\mathbf{z}, \nu_\mu) - 1]\mathbf{m}_{\mu}(\mathbf{z}).
\end{align}\label{eqn:mv_parallel}
Applying the projection operator $\mathcal{P}^\parallel$ on (\ref{eqn:mv}), we have
\begin{align}\label{eqn:mv_long}
\mathcal{P}^\parallel\mathbf{m}^v_{\mu}(\mathbf{r}) =& \int_{\mathbb{R}^3} \D^3z\, \delta^\parallel(\mathbf{r} - \mathbf{z})\tilde{\alpha}(\mathbf{z}, \nu_\mu)\delta(\mathbf{z} - \mathbf{x}_\mu)\mathbf{n}_{j_\mu} \notag \\
&- \int_{\mathbb{R}^3} \D^3 z \, \delta^\parallel(\mathbf{r} - \mathbf{z})\left[\varepsilon(\mathbf{z}, \nu_\mu) - 1\right]\mathbf{m}^v_{\mu}(\mathbf{z}) \notag \\
=& -\frac{\nu^2_\mu}{c^2}\bar{\bar{G}}^\parallel_{0+}(\mathbf{r}, \mathbf{x}_\mu, \nu_\mu) \tilde{\alpha}(\mathbf{x}_\mu, \nu_\mu)\mathbf{n}_{j_\mu} \notag \\
&+ \frac{\nu^2_\mu}{c^2}\int_V \D^3 z\, \bar{\bar{G}}^\parallel_{0+}(\mathbf{r}, \mathbf{z}, \nu_\mu) [\varepsilon(\mathbf{z}, \nu_\mu) - 1]\mathbf{m}_{\mu}(\mathbf{z}),
\end{align}
where we used the expression for the longitudinal delta function (\ref{eqn:delta_long}). From (\ref{eqn:mu+mv_perp}) and (\ref{eqn:mv_long}), using $\mathbf{m}_{\mu}(\mathbf{r}) = \mathcal{P}^\perp\mathbf{m}_{\mu}(\mathbf{r}) + \mathcal{P}^\parallel\mathbf{m}_{\mu}(\mathbf{r})$, $\bar{\bar{G}}_{0+}(\mathbf{r}, \mathbf{z}, \nu) = \bar{\bar{G}}^\perp_{0+}(\mathbf{r}, \mathbf{z}, \nu) + \bar{\bar{G}}^\parallel_{0+}(\mathbf{r}, \mathbf{z}, \nu)$, and (\ref{def:m}), we obtain the integral expression for the $\mathbf{m}$ coefficient (\ref{eqn:m_int}).

\subsection{Differential equations for the e and $\mathbf{m}$ coefficients}
We will show that the solution of the integral equation (\ref{eqn:int_e}) for $\mathbf{e}_\kappa$ satisfies the homogeneous Helmholtz equation. Indeed, after applying the operator $\nabla \times \nabla \times$ on (\ref{eqn:int_e}), using (\ref{eqn:Green_tensor}), one obtains
\begin{align}
    \nabla \times \nabla \times \mathbf{e}_\kappa(\mathbf{r}) =& \frac{\omega_\kappa^2}{c^2}\mathbf{e}_\kappa(\mathbf{r})+ \frac{\omega_\kappa^2}{c^2}\int_V \D^3z[\varepsilon_m(\mathbf{z}, \omega_\kappa)  - 1]\notag \\
&\times \delta(\mathbf{r} - \mathbf{z}) \mathbf{e}_\kappa(\mathbf{z}).
\end{align}
For $\mathbf{x} \in V\backslash\partial V$ (i.e. excluding boundaries of the medium), one obtains 
\begin{equation}
    \nabla \times \nabla \times \mathbf{e}_\kappa(\mathbf{r}) - \frac{\omega_\kappa^2}{c^2}\varepsilon_m(\mathbf{r}, \omega_\kappa)\mathbf{e}_\kappa(\mathbf{r}) = 0,
\end{equation}
and, for $\mathbf{r} \notin V$, the free Helmholtz equation
\begin{equation}
    \nabla \times \nabla \times \mathbf{e}_\kappa(\mathbf{r}) - \frac{\omega_\kappa^2}{c^2}\mathbf{e}_\kappa(\mathbf{r}) = 0, \quad \mathbf{r} \notin V.
\end{equation}
Using the definition of the dielectric coefficient (\ref{def:diel_coef}), we deduce the differential equation for the $\mathbf{e}$ coefficient as 
\begin{equation}
    \nabla \times \nabla \times \mathbf{e}_\kappa(\mathbf{r}) - \frac{\omega_\kappa^2}{c^2}\varepsilon(\mathbf{r}, \omega_\kappa)\mathbf{e}_\kappa(\mathbf{r}) = 0, \quad \mathbf{r} \notin \partial V.
\end{equation} 
We exclude the borders of the dielectric because the dielectric coefficient is discontinuous there.

Similarly, one can apply the differential operator $\nabla \times \nabla \times$ on (\ref{eqn:m_int}) to obtain the differential equation for the $\mathbf{m}$ coefficient for $\mathbf{r} \notin \partial V$
\begin{align}\label{eqn:diff_m}
    \nabla \times& \nabla \times\mathbf{m}_\mu(\mathbf{r}) - \frac{\nu_\mu^2}{c^2}\varepsilon(\mathbf{r}, \nu)\mathbf{m}_\mu(\mathbf{r})   \notag \\ 
    &= -\frac{\nu^2}{c^2}\tilde{\alpha}(\mathbf{x}_\mu, \nu_\mu)\delta(\mathbf{r} - \mathbf{x}_\mu)\mathbf{n}_{j_{_\mu}}.
\end{align}
Here, as for the differential equation for $e_\kappa$, we also exclude the borders $\partial V$ due to the discontinuity of the dielectric coefficient.


\subsection{Solution of the integral equations for $\mathbf{e}_\kappa$ and $\mathbf{m}_\mu$ in terms of the classical Green tensor of the medium}

To establish the connection between a Green tensor of the medium and the operator of the electric field, we introduce the dyadic function $\bar{\bar{G}}_{m+}(\mathbf{r}, \mathbf{x}, \nu)$, as the solution of the following Fredholm integral equation of the second kind
\begin{align}\label{rel:Green}
    \bar{\bar{G}}_{m+}(\mathbf{r}, \mathbf{x}, \nu) =& \bar{\bar{G}}_{0+}(\mathbf{r}, \mathbf{x}, \nu) +  \frac{\nu^2}{c^2}\int_V \D^3z\,\bar{\bar{G}}_{0+}(\mathbf{r}, \mathbf{z}, \nu) \notag \\
   &\times [\varepsilon(\mathbf{z}, \nu) - 1]\bar{\bar{G}}_{m+}(\mathbf{z}, \mathbf{x}, \nu),
\end{align}
where $\mathbf{r},\,\mathbf{x} \in \mathbb{R}^3$. The solution $ \bar{\bar{G}}_{m+}(\mathbf{r}, \mathbf{x}, \nu)$ of this equation is unique \cite{polyanin2008handbook}. Applying the operator $\nabla \times \nabla \times$ on (\ref{rel:Green}) leads to
\begin{equation}\label{eqn:diff_Gm}
	\left( - \frac{\nu^2}{c^2}\varepsilon(\mathbf{r}, \nu) + \nabla \times \nabla \times\right)\bar{\bar{G}}_{m+}(\mathbf{r}, \mathbf{x}, \nu) = \bar{\bar{I}}\delta(\mathbf{r} - \mathbf{x}).
\end{equation}
Moreover, since $\bar{\bar{G}}_{0+}(\mathbf{r}, \mathbf{x}, \nu)$ satisfies the Sommerfeld outgoing radiation condition (see Appendix \ref{app:Green}), (\ref{rel:Green}) implies that $\bar{\bar{G}}_{m+}(\mathbf{r}, \mathbf{x}, \nu)$ also satisfies it: 
\begin{equation}\label{cond:Somm_Gm}
\lim_{|\mathbf{r}| \rightarrow \infty }|\mathbf{r}|\left(\nabla \times - \frac{i\nu}{c}\frac{\mathbf{r}}{|\mathbf{r}|} \times \right)\bar{\bar{G}}_{m+}(\mathbf{r}, \mathbf{x}, \nu) = 0.
\end{equation}
Thus, we conclude that the dyadic tensor $\bar{\bar{G}}_{m+}(\mathbf{r}, \mathbf{x}, \nu)$ is the unique Green tensor for the medium satisfying the Sommerfeld radiation condition. As a consequence \cite[Eq. (E48)]{Semin2024} , $\bar{\bar{G}}_{m+}(\mathbf{r}, \mathbf{x}, \nu)$ satisfies the reciprocity relation $\bar{\bar{G}}_{m+}(\mathbf{x}, \mathbf{r}, \nu) = \bar{\bar{G}}^T_{m+}(\mathbf{r}, \mathbf{x}, \nu)$.

\subsubsection{Solution of the $\mathbf{e}$ coefficients integral equation}

We will show that the solution of Eq.~(\ref{eqn:int_e}) for the $\mathbf{e}$ coefficient,
\begin{align}
\mathbf{e}_\kappa(\mathbf{r}) =&  \omega_{\kappa}\mathbf{\Phi}_\kappa(\mathbf{r}) \notag \\
+& \frac{\omega^2_\kappa}{c^2} \int_V \D^3z\,\bar{\bar{G}}_{0+}(\mathbf{r}, \mathbf{z}, \omega_\kappa) [\varepsilon(\mathbf{z}, \omega_\kappa)  - 1]\mathbf{e}_\kappa(\mathbf{z}),
\end{align}
can  be expressed in terms of the Green tensor of the medium, and of $\mathbf{\Phi}_\kappa$  the transverse continuum eigenfunctions of $\nabla\times\nabla\times$, as
\begin{align}\label{eqn:int_e_2}
	\mathbf{e}_\kappa(\mathbf{r}) =&  \omega_{\kappa}\mathbf{\Phi}_\kappa(\mathbf{r}) +  \frac{\omega^2_\kappa}{c^2} \int_V \D^3z\, \bar{\bar{G}}_{m+}(\mathbf{r}, \mathbf{z}, \omega_\kappa)\notag \\ 
	&\times[\varepsilon(\mathbf{z}, \omega_\kappa) - 1]\omega_{\kappa}\mathbf{\Phi}_\kappa(\mathbf{z}). 
\end{align}
We will prove the identity
\begin{align} \label{prop:e-11}
&\int_{V}\D^3z\,\beta(\mathbf{x})\bar{\bar{G}}_0(\mathbf{r}, \mathbf{z})\mathbf{e}_\kappa(\mathbf{z}) \notag\\
=& \int_{V}\D^3z\,\beta(\mathbf{z})
 \bar{\bar{G}}(\mathbf{r}, \mathbf{z})\, \omega_\kappa\mathbf{\Phi}_\kappa(\mathbf{z}),
\end{align}
with the notation $ \beta(\mathbf{z}) := \frac{\omega^2}{c^2}[\varepsilon( \mathbf{z}, \omega_\kappa) -1]$, $\bar{\bar{G}}_{m+}(\mathbf{r}, \mathbf{z}, \omega_\kappa) = \bar{\bar{G}}(\mathbf{r}, \mathbf{z})$, $\bar{\bar{G}}_{0+}(\mathbf{r}, \mathbf{z}, \omega_\kappa) = \bar{\bar{G}}(\mathbf{r}, \mathbf{z})$, which implies \eqref{eqn:int_e_2}.
\\
The identity (\ref{prop:e-11}) is obtained by replacing $\mathbf{r}$ by $\mathbf{x}$ in Eq.~(\ref{eqn:int_e_2}),   multiplying both sides  by $\beta(\mathbf{r}')\bar{\bar{G}}(\mathbf{r}', \mathbf{x})$ and taking the integral over $\mathbf{x}$:
\begin{align}
& \int_{V}\D^3x\,\beta(\mathbf{x})\bar{\bar{G}}(\mathbf{r}', \mathbf{x})\mathbf{e}_\kappa(\mathbf{x}) \notag\\
&=  \int_{V}\D^3x\,\beta(\mathbf{x})\bar{\bar{G}}(\mathbf{r}', \mathbf{x})\omega_\kappa\mathbf{\Phi}_\kappa(\mathbf{x}) \notag \\
&+  \int_{V}\D^3x\,\beta(\mathbf{x})\int_{V}\D^3z\,\beta(\mathbf{z})\bar{\bar{G}}(\mathbf{r}', \mathbf{x})\bar{\bar{G}}_0(\mathbf{x}, \mathbf{z})\mathbf{e}_\kappa(\mathbf{z}).  \label{B8}
\end{align}
Using (\ref{prop:Green}) the double integral can be be written as
\begin{align}
\int_{V}\D^3z\,\beta(\mathbf{z})
\left[ \int_{V}\D^3x\,\beta(\mathbf{x})\bar{\bar{G}}(\mathbf{r}', \mathbf{x})\bar{\bar{G}}_0(\mathbf{x}, \mathbf{z}) \right]
\mathbf{e}_\kappa(\mathbf{z}) \notag\\
\int_{V}\D^3z\,\beta(\mathbf{z})
\left[\bar{\bar{G}}(\mathbf{r}', \mathbf{z})  - \bar{\bar{G}}_0(\mathbf{r}', \mathbf{z}) \right]
\mathbf{e}_\kappa(\mathbf{z}),
\end{align}
which inserted into \eqref{B8} yields
\begin{align}
& \int_{V}\D^3x\,\beta(\mathbf{x})\bar{\bar{G}}(\mathbf{r}', \mathbf{x})\mathbf{e}_\kappa(\mathbf{x}) 
=  \int_{V}\D^3x\,\beta(\mathbf{x})\bar{\bar{G}}(\mathbf{r}', \mathbf{x})\omega_\kappa\mathbf{\Phi}_\kappa(\mathbf{x}) \notag \\
&+ \int_{V}\D^3z\,\beta(\mathbf{z})
\bar{\bar{G}}(\mathbf{r}', \mathbf{z})  
\mathbf{e}_\kappa(\mathbf{z}) 
-\int_{V}\D^3z\,\beta(\mathbf{z})
   \bar{\bar{G}}_0(\mathbf{r}', \mathbf{z}) 
\mathbf{e}_\kappa(\mathbf{z}).
\end{align}
The second term is equal to the left hand side, and thus it
implies \eqref{prop:e-11}.

\subsubsection{Solution of the $\mathbf{m}$ coefficients integral equation}

Furthermore, we next show that the solution of Eq.~(\ref{eqn:m_int}) for the $\mathbf{m}$ coefficient 
\begin{align}
\mathbf{m}_{\mu}(\mathbf{r}) &=  - \tilde{\alpha}( \mathbf{r}, \nu_\mu)\frac{\nu^2}{c^2} \bar{\bar{G}}_{0+}(\mathbf{r}, \mathbf{x}_\mu, \nu_\mu)\mathbf{n}_{j_\mu} \notag \\
&+ \frac{\nu_\mu^2}{c^2}\int_V \D^3z\,\bar{\bar{G}}_{0+}(\mathbf{r}, \mathbf{z}, \nu) [\varepsilon(\mathbf{z}, \nu_\mu) - 1]\mathbf{m}_{\mu}(\mathbf{z}),
\end{align}
can also be expressed in terms of the Green tensor $\bar{\bar{G}}_{m+}(\mathbf{r}, \mathbf{x}, \omega_\kappa)$ of the medium:
\begin{equation}\label{eqn:m_green}
	\mathbf{m}_{\mathbf{x}, \nu, j}(\mathbf{r}) = -\tilde{\alpha}(\mathbf{x}, \nu)\frac{\nu^2}{c^2}\bar{\bar{G}}_{m+}(\mathbf{r}, \mathbf{x}, \nu) \cdot \mathbf{n}_j.
\end{equation}
Applying the vector $\mathbf{b} := - \tilde{\alpha}(\mathbf{x}, \nu)\frac{\nu^2}{c^2}\mathbf{n}_j$ on the right side
of Eq.~(\ref{rel:Green}) we obtain
\begin{align}
\bar{\bar{G}}_{m+}(\mathbf{r}, \mathbf{x}, \nu)\cdot \mathbf{b} =& \bar{\bar{G}}_{0+}(\mathbf{r}, \mathbf{x}, \nu)\cdot \mathbf{b} + \int_V \D^3z\,\bar{\bar{G}}_{0+}(\mathbf{r}, \mathbf{z}, \nu) \notag \\
&\times \frac{\nu^2}{c^2}[\varepsilon(\mathbf{z}, \nu) - 1]\bar{\bar{G}}_{m+}(\mathbf{z}, \mathbf{x}, \nu) \cdot \mathbf{b},
\end{align}
which shows that $\bar{\bar{G}}_{m+}(\mathbf{r}, \mathbf{x}, \nu)\cdot \mathbf{b}$ satisfies the Eq.~(\ref{eqn:m_int}). Since (\ref{eqn:m_int}) is a Fredholm equation with a unique solution, we conclude that
\begin{equation}
	\mathbf{m}_{\mathbf{x},\nu, j}(\mathbf{r}) = -\tilde{\alpha}(\mathbf{x}, \nu)\frac{\nu^2}{c^2}\bar{\bar{G}}_{m+}(\mathbf{r}, \mathbf{x}, \nu) \cdot \mathbf{n}_j.
\end{equation}
Thus, it is sufficient to determine the dyadic Green tensor from the integral equation (\ref{rel:Green}) or from (\ref{eqn:diff_Gm}) and (\ref{cond:Somm_Gm}) to obtain the complete expression of the electric field operator.

\subsection{Operator of the electric field observable}
The final form of the operator of the electric field is 
\begin{equation}\label{electric_field_op}
\hat{\mathbf{E}}(\mathbf{r}) = \hat{\mathbf{E}}^e(\mathbf{r}) + \hat{\mathbf{E}}^m(\mathbf{r}),
\end{equation}
where
\begin{align}
\hat{\mathbf{E}}^e(\mathbf{r}) &= -\int \D\kappa \sqrt{\frac{\hbar}{2\varepsilon_0 \omega_\kappa }}\left[ \mathbf{e}_\kappa(\mathbf{r}) \hat{C}^e_\kappa + \text{H.c.}\right],\\
 \hat{\mathbf{E}}^m(\mathbf{r}) &= -\int \D\mu \sqrt{\frac{\hbar}{2\varepsilon_0 \nu_\mu }}\left[ \mathbf{m}_\mu(\mathbf{r}) \hat{C}^m_\mu + \text{H.c.}\right].
 \end{align}
To make the link with the expression reported in the literature, using (\ref{eqn:m_green}), we can rewrite the medium part of the electric field $ \hat{\mathbf{E}}^m(\mathbf{r})$ in the following form
\begin{equation}
\hat{\mathbf{E}}^m(\mathbf{r}) = \int \D\nu \int_V \D^3x\, \left[i\nu\bar{\bar{G}}_{m+}(\mathbf{r}, \mathbf{x}, \nu) \hat{\mathbf{j}}(\mathbf{x}, \nu) + \text{H.c.}\right],
\end{equation}
where we introduced the current operator
\begin{equation}
\hat{\mathbf{j}}(\mathbf{x}, \nu) = -i\frac{\nu}{c^2}\sqrt{\frac{\hbar}{\pi \varepsilon_0}\varepsilon(\mathbf{x}, \nu)}\sum_j \mathbf{n}_j\hat{C}^m_{\mathbf{x}, \nu, j}.
\end{equation}
Thus, the contribution from the medium $\hat{\mathbf{E}}^m(\mathbf{r})$ for the electric field has the same form as the one obtained for a bulk medium \cite{Gruner1996, Dung1998three, Scheel1998}. The term $\hat{\mathbf{E}}^e(\mathbf{r})$ has to be added for the case of a finite medium.
\section {Properties of field coefficients}\label{sec:Prop}

\subsection{Eigenfunctions of the Lippmann-Schwinger equations expressed in terms of the medium Green function
\label{V-A}}

In this section, we show that  the continuum eigenfunctions $\underline{\psi}^e_\kappa = \begin{pmatrix} u^e_{\kappa}(\kappa') \\ v^e_{\kappa}(\mu') \end{pmatrix}$ and $\underline{\psi}^m_\mu = \begin{pmatrix} u^m_{\mu}(\kappa') \\ v^m_{\mu}(\mu') \end{pmatrix}$ of $\Omega^2$, solutions of the Lippmann-Schwinger equations
can also be written explicitly in terms of the $e$ and $\mathbf{m}$ coefficients, and therefore also in terms of the classical Green tensor $\bar{\bar{G}}_{m+}(\mathbf{r}, \mathbf{x}, \nu)$ of the medium. One can explicitly express the eigenfunction directly from (\ref{eqn:LS_e}) and (\ref{eqn:LS_m}) rewriting the integral operators as follows
\begin{align}
[B_1 \mathbf{v}^e_\kappa ](\kappa') & = \int_{\mathbb{R}^3} \D^3r\, \omega_{\kappa'}\mathbf{\Phi}_{\kappa'}(\mathbf{r}) \cdot \mathcal{P}^\perp\mathbf{e}^v_\kappa(\mathbf{r}),\\
[B_2 u^e_{\kappa}](\mu') &= \tilde{\alpha}(\mathbf{x}', \nu')\mathbf{n}_{j'} \cdot \mathbf{e}^u_{\kappa'}(\mathbf{r}),\\
[A\mathbf{v}^e_{\kappa}](\mu') &= \tilde{\alpha}(\mathbf{x}', \nu') \mathbf{n}_{j'} \cdot \mathbf{e}^v_{\kappa'}(\mathbf{r})
\end{align}
for the eigenfunctions $\underline{\psi}^e_\kappa$ related to the free electromagnetic field, and
\begin{align}
[B_1 \mathbf{v}^m_{\mu} ](\kappa') & = \int_{\mathbb{R}^3} \D^3r\, \omega_{\kappa'} \mathbf{\Phi}_{\kappa'}(\mathbf{r}) \cdot \mathcal{P}^\perp \mathbf{m}^v_\mu(\mathbf{r}),\\
[B_2 u^m_{\mu}](\mu') &= \tilde{\alpha}(\mathbf{x}', \nu')\mathbf{n}_{j'} \cdot\mathbf{m}^v_{\mu}(\mathbf{r}),\\
[A\mathbf{v}^m_{\mu}](\mu') &= \tilde{\alpha}(\mathbf{x}', \nu')\mathbf{n}_{j'} \cdot\mathbf{m}^v_{\mu}(\mathbf{r})
\end{align}
for the eigenfunctions $\underline{\psi}^m_\mu$ related the medium. Then, one can substitute the integral operators to (\ref{eqn:LS_e}) and (\ref{eqn:LS_m}) and obtain 
\begin{subequations}\label{eqn:eig_e}
\begin{align}
&u^e_{\kappa}(\kappa') = \delta(\kappa - \kappa') - \frac{\int_V \D^3r\, \omega_\kappa \mathbf{\Phi}_\kappa(\mathbf{r}) \mathcal{P}^\perp\mathbf{e}^v_\kappa(\mathbf{r})}{\omega_{\kappa'}^2 - \omega_{\kappa}^2 - i0} \label{eqn:LS_e_u}\\
&\mathbf{v}^e_{\kappa}(\mathbf{x}', \nu')  = - \frac{\tilde{\alpha}(\mathbf{x}', \nu') \mathbf{e}_{\kappa}(\mathbf{x}')}{\nu'^2 - \omega_{\kappa}^2 - i0}, \label{eqn:LS_e_v}
\end{align}
\end{subequations}
for the deformed modes of the electromagnetic field and
\begin{subequations}\label{eqn:eig_m}
\begin{align}
&u^m_{\mu}(\kappa') =  - \frac{\int_V \D^3r\, \omega_{\kappa'} \mathbf{\Phi}_{\kappa'}(\mathbf{r})  \mathcal{P}^\perp\mathbf{m}^v_\mu(\mathbf{r})}{\omega_{\kappa'}^2 - \nu_{\mu}^2 - i0} \\
&\mathbf{v}^m_{\mu}(\mathbf{x}', \nu')  = \mathbf{n}_{j}\delta(\mathbf{x} - \mathbf{x}')\delta(\nu - \nu') - \frac{\tilde{\alpha}(\mathbf{x}', \nu')  \mathbf{m}_{\mu}(\mathbf{x})}{\nu'^2 - \nu_{\mu}^2 - i0}. 
\end{align}
\end{subequations}
for the deformed modes of the medium.

\subsection{Green tensor LDOS identity}
Another practical relation one can obtain is the imaginary green tensor identity, which connects the Green tensor of the field coupled with the medium and the $e$- and $\mathbf{m}$ coefficients. Using the connection between Green tensors (\ref{rel:Green}), the integral relation on the $\mathbf{e}$ coefficient and the Green tensor (\ref{eqn:int_e_2}), and 
\begin{align} \label{ImG0-Phi}
&\mathrm{Im}\, \bar{\bar{G}}_{0}(\mathbf{x}, \mathbf{y}, \omega)  =
\mathrm{Im}\, \bar{\bar{G}}_{0}^\perp(\mathbf{x}, \mathbf{y}, \omega) \notag\\
&= \frac{\pi c^2}{2 \omega} 
\int d\kappa~ \mathbf{\Phi}_\kappa(\mathbf{x}) \otimes \mathbf{\Phi}_{\kappa}(\mathbf{y}) \delta(\omega_\kappa-\omega),
\end{align}
where $\mathbf{\Phi}_{\kappa}$ are the transverse continuum eigenfunctions \eqref{def-Phi_kappa}  of $\nabla\times\nabla\times$, 
one obtains the relation for the Green tensor of the field coupled with the dielectric
\begin{align}\label{eqn:IGTI}
    \mathrm{Im}[&\bar{\bar{G}}_{m+}(\mathbf{x}, \mathbf{y}, \omega)] = \frac{\pi c^2}{2 \omega^3}\int \D \kappa\, \mathbf{e}_{\kappa}(\mathbf{x}) \otimes \mathbf{e}^*_{\kappa}(\mathbf{y})\delta(\omega - \omega_\kappa) \notag \\
    &+ \frac{\omega^2}{c^2} \int \D^3 z\, \varepsilon_i(\mathbf{z}, \omega) \bar{\bar{G}}_{m+}(\mathbf{x},\mathbf{z}, \omega) \bar{\bar{G}}_{m+}^*(\mathbf{z}, \mathbf{y}, \omega).
\end{align}
A complete derivation is given in Appendix \ref{app:Proof}. One can apply the relation between the Green tensor of the medium and the $\mathbf{m}$ coefficient (\ref{eqn:m_green}) to express the Green tensor LDOS identity in terms of the $e$ and $\mathbf{m}$ coefficients:
\begin{align}\label{LDOS_em}
	 \mathrm{Im}[&\bar{\bar{G}}_{m+}(\mathbf{x}, \mathbf{y}, \omega)] = \frac{\pi c^2}{2 \omega^3}\int \D \kappa\, \mathbf{e}_{\kappa}(\mathbf{x}) \otimes \mathbf{e}^*_{\kappa}(\mathbf{y})\delta(\omega - \omega_\kappa) \notag \\
 &+ \frac{\pi c^2}{2 \omega^3} \int \D\mu \, \mathbf{m}_{\mu}(\mathbf{x}) \otimes \mathbf{m}^*_{\mu}(\mathbf{y})\delta(\omega - \nu_\mu).
\end{align}

\subsection{Uncoupling limit}
Without the medium, the electromagnetic field should behave as a free field. To validate the results in such a case, we make the field "blind" to the matter by turning off the interaction between them and considering the limit $\tilde{\alpha}\rightarrow 0$ or $\varepsilon_i \rightarrow 0$, which leads to $\varepsilon \rightarrow 1$ due to the Kramers-Kronig relation. We call such a limit as \textit{uncoupling limit} and mark it as $\lim_{\text{nc}} = \lim_{\tilde{\alpha}\rightarrow 0} = \lim_{\varepsilon \rightarrow 1}$.

In the uncoupling limit, we can immediately obtain the results for the field coefficients. First, we consider the $\mathbf{e}$ coefficients. From (\ref{eqn:evk}) we have 
\begin{equation}
\lim_{\text{nc}}\mathbf{e}^v_\kappa (\mathbf{x}) = 0,
\end{equation}
which implies $\mathbf{e}_\kappa (\mathbf{x}) = \mathbf{e}^u_\kappa (\mathbf{x})$. Applying the limit to (\ref{eqn:int_e}) yields
\begin{equation}\label{lim:e}
\lim_{\text{nc}}\mathbf{e}_\kappa (\mathbf{x}) =  \lim_{\text{nc}}\mathbf{e}^u_\kappa (\mathbf{x}) = \omega_\kappa \mathbf{\Phi}(\mathbf{x}).
\end{equation}
Thus, in the $\mathbf{e}$ coefficients, only the term associated with the free electromagnetic field remains. On the other hand, the $\mathbf{m}$ coefficients are zero in the uncoupling limit. Indeed, by applying the limit on (\ref{eqn:mu}) and (\ref{eqn:mv}), we have
\begin{equation}
\lim_{\text{nc}}\mathbf{m}^v_\mu (\mathbf{x}) = \lim_{\text{nc}}\mathbf{m}^u_\mu (\mathbf{x}) = 0.
\end{equation}
Thus, the medium part of the electromagnetic operator disappears
$\lim_{\text{nc}} \hat{\mathbf{E}}^m = 0$, and
\begin{equation}
\lim_{\text{nc}} \hat{\mathbf{E}} = \hat{\mathbf{E}}^{free}.
\end{equation}
The Green tensor of the field coupled with the medium should also turn into the Green tensor of the free field in the uncoupling limit. Indeed, taking the limit of Eq.~(\ref{rel:Green}), we have 
\begin{equation}
\lim_{\text{nc}} \bar{\bar{G}}_{m+}(\mathbf{x}, \mathbf{y}, \omega) =\bar{\bar{G}}_{0+}(\mathbf{x}, \mathbf{y}, \omega),
\end{equation}
since the integral term disappears. The same can be done with the Green tensor LDOS identity,
\begin{align}
    \lim_{\text{nc}}&\, \mathrm{Im}[\bar{\bar{G}}_{m+}(\mathbf{x}, \mathbf{y}, \omega)] = \lim_{\text{nc}}\left[\frac{\pi c^2}{2 \omega^3}\sum_{\sigma} \mathbf{e}_{\omega, \sigma}(\mathbf{x}) \otimes \mathbf{e}^*_{\omega,\sigma}(\mathbf{y})\right. \notag \\
    &\left.+ \frac{\omega^2}{c^2} \int \D^3 z\, \varepsilon_i(\mathbf{z}, \omega) \bar{\bar{G}}_{m+}(\mathbf{x},\mathbf{z}, \omega) \bar{\bar{G}}_{m+}^*(\mathbf{z}, \mathbf{y}, \omega)\right]. \notag \\
    &= \frac{\pi c^2}{2 \omega}\sum_{\sigma} \mathbf{\Phi}_\kappa(\mathbf{x}) \otimes \mathbf{\Phi}_\kappa(\mathbf{y}) = \mathrm{Im}[ \bar{\bar{G}}_{0+}(\mathbf{x}, \mathbf{y}, \omega)],
\end{align}
where we used (\ref{lim:e}), and the integral term disappears because it contains the imaginary part of the dielectric coefficient.

The eigenfunctions $\underline\psi^e_\kappa$ and $\underline\psi^m_\mu$ take the form of $\underline\phi^e_\kappa$ and $\underline\phi^m_\mu$ in the uncoupling limit. One can see clearly from (\ref{eqn:eig_e}) and (\ref{eqn:eig_m}), that
\begin{subequations}
\begin{align}
\lim_{\text{nc}}u^e_{\kappa}(\kappa') &= \delta(\kappa - \kappa'), \\
\lim_{\text{nc}} \mathbf{v}^e_{\kappa}(\mathbf{x}', \nu') &= 0, \\
\lim_{\text{nc}}u^m_{\mu}(\kappa') &= 0, \\
\lim_{\text{nc}} \mathbf{v}^m_{\mu}(\mathbf{x}', \nu')  &= \mathbf{n}_{j}\delta(\mathbf{x} - \mathbf{x}')\delta(\nu - \nu'),
\end{align}
\end{subequations}
which agrees with the fact that the Hamiltonian in this limit is already in diagonal form. Thus, all the crucial properties of the field have the form of a free field in the uncoupling limit.  

\section{Purcell effect}

We consider the model of a two-level quantum emitter of states $|g\rangle$, $|e\rangle$ with the ground and excited energy 0 and $\hbar\omega_{a}$, respectively, located at $\mathbf{r}_a$, and coupled with the QPP field in the rotating-wave approximation \cite{dzsotjan2010quantum}. The spontaneous emission rate of the quantum emitter is given by Fermi's Golden rule
\cite[Sect. 8.4.1, Eq. (8.89)]{novotny2012principles}
\begin{equation}
\Gamma(\mathbf{r}_a, \omega_a) = \frac{2\pi}{\hbar^2} \int \D\eta\, \left|\langle g \otimes 1_\eta| \hat{W}(\omega_a) | e \otimes \varnothing \rangle \right|^2 \delta(\omega_a - \omega_\eta),
\end{equation} 
where $\hat{W} = - \left[\hat{\sigma}_{eg} \otimes \mathbf{d}\cdot\hat{\mathbf{E}}(\mathbf{r}_a) + \text{H.c.}\right]$ is the interaction Hamiltonian written in the rotation-wave approximation with the dipole moment $d$ and the emitter operator $\sigma_{eg} = |e\rangle\langle g|$, and $\eta$ labels all the possible one-QPP states $|1_\eta \rangle$, of energy $\hbar \omega_\eta$, associated with the decay from the initial excited state $|e\rangle$ (and the vacuum for the QPP) state to the ground state $|g\rangle$. Substituting the electric field operator expression (\ref{electric_field_op}),  which includes the contributions of the two continua,
\begin{align}  
\hat{\mathbf{E}}(\mathbf{r}) &= \hat{\mathbf{E}}^e(\mathbf{r}) + \hat{\mathbf{E}}^m(\mathbf{r}),\\  \label{e-term-in-E}
\hat{\mathbf{E}}^e(\mathbf{r}) &= -\int \D\kappa \sqrt{\frac{\hbar}{2\varepsilon_0 \omega_\kappa }}\left[ \mathbf{e}_\kappa(\mathbf{r}) \hat{C}^e_\kappa + \text{H.c.}\right]\\
 \hat{\mathbf{E}}^m(\mathbf{r}) &= -\int \D\mu \sqrt{\frac{\hbar}{2\varepsilon_0 \nu_\mu }}\left[ \mathbf{m}_\mu(\mathbf{r}) \hat{C}^m_\mu + \text{H.c.}\right],
 \end{align}
we can write, after applying the creation-annihilation operators to the initial state, evaluating the scalar products and integrating over the final states $d\eta$, the decay rate in the following form 
\begin{equation}
\Gamma(\mathbf{r}_a, \omega_a) = \Gamma_e(\mathbf{r}_a, \omega_a) + \Gamma_m(\mathbf{r}_a, \omega_a),
\end{equation} 
with
\begin{subequations}
\begin{align}
\Gamma_e(\mathbf{r}_a, \omega_a) = \frac{\pi}{\hbar\omega\varepsilon_0}\int& \D\kappa\, \delta(\omega_\kappa - \omega_a)\notag \\ 
&\times \mathbf{d}\cdot \mathbf{e}_{\kappa}(\mathbf{r}_a) \otimes \mathbf{e}^*_{\kappa}(\mathbf{r}_a)  \cdot \mathbf{d},
\label{Gae-term}
\end{align}
\begin{align}
\Gamma_m(\mathbf{r}_a, \omega_a) = \frac{\pi}{\hbar\omega\varepsilon_0}\int& \D\mu\, \delta(\nu_\mu - \omega_a) \notag \\
&\times \mathbf{d}\cdot \mathbf{m}_{\mu}(\mathbf{r}_a) \otimes \mathbf{m}^*_{\mu}(\mathbf{r}_a)  \cdot \mathbf{d}.
\label{Gam-term}
\end{align}
\end{subequations}
Rewriting the term $\Gamma_m(\mathbf{r}_a, \omega_a)$ using \eqref{eqn:m_green}, 
\begin{equation}\label{eqn:m_green-22}
	\mathbf{m}_{\mathbf{x}, \nu, j}(\mathbf{r}) = -\tilde{\alpha}(\mathbf{x}, \nu)\frac{\nu^2}{c^2}\bar{\bar{G}}_{m+}(\mathbf{r}, \mathbf{x}, \nu) \cdot \mathbf{n}_j,
\end{equation}
and the  Green tensor LDOS identity (\ref{LDOS_em}),
\begin{align}\label{LDOS_em-22}
	 \mathrm{Im}[&\bar{\bar{G}}_{m+}(\mathbf{x}, \mathbf{y}, \omega)] = \frac{\pi c^2}{2 \omega^3}\int \D \kappa\, \mathbf{e}_{\kappa}(\mathbf{x}) \otimes \mathbf{e}^*_{\kappa}(\mathbf{y})\delta(\omega - \omega_\kappa) \notag \\
  &+ \frac{\pi c^2}{2 \omega^3} \int \D\mu \, \mathbf{m}_{\mu}(\mathbf{x}) \otimes \mathbf{m}^*_{\mu}(\mathbf{y})\delta(\omega - \nu_\mu),
\end{align}
 we obtain
\begin{align}\label{eqn:Gamma_m}
&\Gamma_m(\mathbf{r}_a, \omega_a) = \frac{2\omega_a^2}{\hbar\varepsilon_0c^2} \mathbf{d} \cdot \mathrm{Im}[\bar{\bar{G}}_{m+}(\mathbf{r}_a, \mathbf{r}_a, \omega)] \cdot \mathbf{d}\notag \\
& - \frac{\pi}{\hbar\omega\varepsilon_0} 
\int \D\kappa\, \delta(\omega-\omega_\kappa)\mathbf{d}\cdot \mathbf{e}_{\kappa}(\mathbf{r}_a) \otimes \mathbf{e}^*_{\kappa}(\mathbf{r}_a) \cdot \mathbf{d}.
\end{align}
Summing $\Gamma_e$ and $\Gamma_m$, the term with the $\mathbf{e}$ coefficient in (\ref{eqn:Gamma_m}) compensates exactly the contribution \eqref{Gae-term} related to free electric field $\Gamma_e$, and we obtain a simple expression for the decay rate 
%
\begin{align} \label{Ga_final}
\Gamma(\mathbf{r}_a, \omega_a) = 
 \frac{2\omega_a^2}{\hbar\varepsilon_0c^2} \mathbf{d} \cdot \mathrm{Im}[\bar{\bar{G}}_{m+}(\mathbf{r}_a, \mathbf{r}_a, \omega)] \cdot \mathbf{d}.
\end{align} 
Thus, the Purcell factor, defined as $P(\mathbf{r}_a, \omega_a) = \Gamma(\mathbf{r}_a, \omega_a) / \Gamma_0(\mathbf{r}_a, \omega_a)$ with the decay rate in the vacuum $\Gamma_0(\mathbf{r}_a, \omega_a) = \omega^3 |\mathbf{d}|^2 / (3\pi \varepsilon_0 \hbar c^3)$, reads as 
\begin{equation} \label{final-Purcell-hrj}
P(\mathbf{r}_a, \omega_a) = \frac{2\pi c }{\omega_0|\mathbf{d}|^2}\mathbf{d} \cdot \mathrm{Im}\left[\bar{\bar{G}}_{m+}(\mathbf{r}_a, \mathbf{r}_a, \omega_a)\right] \cdot \mathbf{d}.
\end{equation}
This expression coincides with the formula obtained in the literature \cite{novotny2012principles} using the formulas for a bulk. 
\\ \\
{\bf Remarks:} The exact compensation of the terms involving the $\mathbf{e}$ coefficients in \eqref{eqn:Gamma_m}, that come from the LDOS identity \eqref{LDOS_em-22}, with those of \eqref{Gae-term}, provide an explanation of why the Purcell factor for a finite medium involving a double continuum coincides with the one obtained for a bulk medium that involves a single continuum. In particular, it explains why the result coincides with the approach described e.g. in \cite{buhmann2013dispersion, Drezet2017b, Hanson2021}, consisting of adding a small dissipative dielectric coefficient extending to infinity to the one for the finite medium, which eliminates the e-terms in the LDOS identity \eqref{LDOS_em-22} and in  \eqref{Gae-term}, since the $\hat{\mathbf{E}}^e(\mathbf{r})$  contribution \eqref{e-term-in-E} to the electric field observable disappears. In this approach, one sets the small dielectric coefficient to zero at the end of the calculation. The fact that the two approaches lead to the same Purcell factor is explained by the fact that the terms that are made to disappear by adding the small dissipative bulk term are exactly the ones that compensate each other exactly in the model for a finite medium without the infinite bulk addition. We emphasize that the decay rate is driven by the coupling with the two terms $\mathbf{E}^e$ and $\mathbf{E}^m$ of the electric field and not just by the medium one $\mathbf{E}^m$ since if it were so the decay rate would be (\ref{eqn:Gamma_m}) and not (\ref{Ga_final}).


\section{Conclusion and outlook}\label{sec:Conclusion}

This work extends the earlier results that were formulated for a one-dimensional quantum plasmonic system \cite{Semin2024} to three-dimensional systems with a finite medium of arbitrary shape. In contrast to the one-dimensional system, the electric field in three dimensions has transversal and longitudinal components due to the additional degrees of freedom associated with the oscillations in three dimensions of the charges of the medium. 

The main result is that all the elements of the quantized plasmonics fields can be expressed in terms of the classical Green tensor of the medium, for which there are well-established numerical algorithms. This approach can be extended in a straightforward way to anisotropic and to magnetic media. 

In future works, the presented approach can be applied to other physical phenomena, e.g., superradiance, Casimir effect, or quantum control of emitters in nanostructured environments. Another direction is the construction of simplified reduced models involving only a small number of dominating modes \cite{buhmann2008casimir, hummer2013weak, sauvan2013theory, Rousseaux2016, Dzsotjan2016, Castellini2018, franke2019quantization, medina2021few, sanchez2022few, lednev2024spatiallyresolvedphotonstatistics}.

\section{Acknowledgments}
We acknowledge support from the EUR-EIPHI Graduate School (Grant No. 17-EURE-0002) and the QuanTEdu- France project (Grant No. ANR-22-CMAS-0001). We thank G\'erard Colas des Francs and Jonas Lampart for many fruitful exchanges during the preparation of this work.

\appendix

\section{Green tensor of the free electromagnetic field}\label{app:Green}

The Green tensor of the free electromagnetic field is the solution of the Helmholtz equation
\begin{equation}
\left[\nabla \times \nabla \times -\frac{\omega^2}{c^2}\mathbb{I}\right] \bar{\bar{G}}_{0+}(\mathbf{r}, \mathbf{r}', \omega) = \bar{\bar{I}}\delta(\mathbf{r}-\mathbf{r}'). 
\end{equation}
The unique solution associated with the outgoing wave must satisfy the Sommerfeld radiation condition of the following form \cite{sommerfeld1949partial}
\begin{equation}\label{lim:Somm_cond_tensor}
\lim_{r\rightarrow \infty} r\left(\nabla \times   - i\frac{\omega}{c}\frac{\mathbf{r}}{r} \times \right)\bar{\bar{G}}_{0+}(\mathbf{r}, \mathbf{r}', \omega) = 0.
\end{equation}
One can express the Green tensor of the free electromagnetic field as 
\begin{equation}\label{eqn:Green_tensor_function}
\bar{\bar{G}}_{0+}(\mathbf{r}, \mathbf{r}', \omega) = \left(\bar{\bar{I}} + \frac{\nabla \otimes \nabla}{{\omega^2}/{c^2}} \right)g(\mathbf{r}, \mathbf{r}', \omega),
\end{equation}
where $g(\mathbf{r}, \mathbf{r}', \omega)$ is the scalar Green function of the scalar Laplacian for the free field
\begin{equation}
g(\mathbf{r}, \mathbf{r}', \omega) = \frac{1}{4\pi}\frac{e^{i\frac{\omega}{c}|\mathbf{r} - \mathbf{r}'|}}{|\mathbf{r} - \mathbf{r}'|},
\end{equation}
that satisfies the scalar Sommerfeld outgoing radiation condition
\begin{equation}
\lim_{\mathbf{r}\rightarrow \infty}r\left( \frac{\partial}{\partial r}- i\frac{\omega}{c}\right)g(\mathbf{r}, \mathbf{r}', \omega)  = 0.
\end{equation}  
The evaluation of the derivatives in (\ref{eqn:Green_tensor_function}), together with an analysis of the singularity at $\mathbf{r}=\mathbf{r}'$ leads to the following explicit formula \cite{Knoll2003, buhmann2013dispersion}
\begin{align}
\bar{\bar{G}}_{0+}(\mathbf{r}, \mathbf{r}', \omega) & = -\frac{c^2}{3\omega^2}\delta(\mathbf{r} - \mathbf{r}')\bar{\bar{I}}  \notag \\ 
&+ \left[\left( \frac{3c^2}{\omega^2R^2} - \frac{3ic}{\omega R} - 1\right)\hat{\mathbf{R}}\otimes\hat{\mathbf{R}} \right. \notag \\
&+ \left.\left(1 + \frac{ic}{\omega R} - \frac{c^2}{\omega^2R^2}\right)\bar{\bar{I}} \right]g(\mathbf{r}, \mathbf{r}', \omega),
\end{align}
where $R = |\mathbf{r} - \mathbf{r}'|$ and $\hat{\mathbf{R}} = (\mathbf{r} - \mathbf{r}')/R$.


Another way is to write the free Green tensor in terms of the continuum eigenfunctions $\mathbf{\Phi}_{\kappa}(\mathbf{r})$ for the double curl operator $\nabla \times \nabla \times$
\begin{equation}\label{eqn:Green_eigen}
\bar{\bar{G}}_{0\pm}(\mathbf{r}, \mathbf{r}', \omega) = \int \D\tilde{\kappa}  \frac{\mathbf{\Phi}_{\tilde{\kappa}}(\mathbf{r}) \otimes \mathbf{\Phi}_{\tilde{\kappa}}(\mathbf{r}')}{\lambda^2_{\tilde\kappa}- \omega^2/c^2 \mp i0^+},
\end{equation}
where $\tilde{\kappa} = (\mathbf{k}, \sigma, \zeta)$, $\mathbf{k} \in \mathbb{R}^2 \times [0, \infty)$, $\sigma \in \{0, \pm\}$, $\zeta \in \{c, s\}$, 
\begin{equation}
\nabla \times \nabla \times \mathbf{\Phi}_{\tilde{\kappa}}(\mathbf{r}) = \lambda^2_{\tilde\kappa} \mathbf{\Phi}_{\tilde{\kappa}}(\mathbf{r}),
\end{equation}
with the completeness given by
\begin{equation}
\int\D\kappa\, \mathbf{\Phi}_{\tilde{\kappa}}(\mathbf{r}) \otimes \mathbf{\Phi}_{\tilde{\kappa}}(\mathbf{r}') = \bar{\bar{I}}\delta(\mathbf{r} - \mathbf{r}').
\end{equation}
Comparing to Sec.\ref{sec:canon_var}, here we consider the full set of eigenvectors including the longitudinal components
\begin{equation}
\mathbf{\Phi}_{\mathbf{k}, 0, \zeta}(\mathbf{r}) = (2\pi)^{-3/2}\vec{\epsilon}_0(\mathbf{k})
\begin{cases}
\cos(\mathbf{k}\mathbf{r}), & \zeta = c, \\
\sin(\mathbf{k}\mathbf{r}), & \zeta = s,
\end{cases}
\end{equation}
where $\vec{\epsilon}_0(\mathbf{k}) = \mathbf{k}/|\mathbf{k}|$. The longitudinal eigenfunctions constitute the kernel of the double curl operator
\begin{equation}
\nabla \times \nabla \times \mathbf{\Phi}_{\mathbf{k}, 0, \zeta}(\mathbf{r}) = 0.
\end{equation}
Choosing the sign $-i0^+$ or $+i0^+$ in the denominator, the Green tensor satisfies the Sommerfeld radiation condition for either the outgoing or the incoming wave, respectively. We work with the outgoing wave and denote the corresponding Green function by $\bar{\bar{G}}_{0+}(\mathbf{r}, \mathbf{r}', \omega)$. 

We can rewrite it as a sum of transverse and longitudinal terms as follows
\begin{equation}
\bar{\bar{G}}_{0+}(\mathbf{r}, \mathbf{r}', \omega) = \bar{\bar{G}}_{0+}^\perp(\mathbf{r}, \mathbf{r}', \omega) + \bar{\bar{G}}_{0+}^\parallel(\mathbf{r}, \mathbf{r}', \omega).
\end{equation}
The transverse term $\bar{\bar{G}}_{0+}^\perp(\mathbf{r}, \mathbf{r}', \omega)$ is constructed from the transverse eigenvectors (i.e. $\sigma = \pm$). It satisfies the following differential equation
\begin{equation}
\nabla \times \nabla \times\bar{\bar{G}}_{0+}^\perp(\mathbf{r}, \mathbf{r}', \omega) - \frac{\omega^2}{c^2}\bar{\bar{G}}_{0+}^\perp(\mathbf{r}, \mathbf{r}', \omega) = \delta^\perp(\mathbf{r} -\mathbf{r}'),
\end{equation}
where $\delta^\perp(\mathbf{r} - \mathbf{r}')$ is the transverse delta-function.
The longitudinal term $\bar{\bar{G}}_{0}^\parallel(\mathbf{r}, \mathbf{r}', \omega)$ is constructed from the longitudinal eigenvectors with $\sigma = 0$, and it satisfies the condition for the longitudinal eigenvectors
\begin{equation}
\nabla \times \nabla \times\bar{\bar{G}}_{0+}^\parallel(\mathbf{r}, \mathbf{r}', \omega) = 0.
\end{equation}
Thus, we can immediately find the expression of the longitudinal term of the dyadic Green tensor for the free field from the Helmholtz equation
\begin{equation}
- \frac{\omega^2}{c^2}\bar{\bar{G}}_{0+}^\parallel(\mathbf{r}, \mathbf{r}', \omega) = \delta^\parallel(\mathbf{r} - \mathbf{r}')
\end{equation}
with the longitudinal delta-function $\delta^\parallel(\mathbf{r} - \mathbf{r}')$.

\section{Proof of the integral relations between $\bar{\bar{G}}_{m+}$ and $\bar{\bar{G}}_{0+}$ 
and the $\mathbf{e}$ coefficients \label{app:Proof}  }
In this appendix, we derive some identities involving the Green tensors  $\bar{\bar{G}}_{m+}$ and $\bar{\bar{G}}_{0+}$ ,
and we prove the expression of the solution of the integral equation for the $\mathbf{e}$ coefficients in terms of the medium Green tensor. We use these results to prove the Green tensor LDOS identity for a finite medium, 
which plays a central role in the  treatment of spontaneous emission. We will use the following notation:
\begin{subequations}
\begin{equation}
\beta(\mathbf{z}, \omega) =\frac{\omega^2}{c^2}\left[ \varepsilon_m(\mathbf{z},\omega) - 1\right],
\end{equation}
\begin{equation}
\bar{\bar{G}}_{0+} \equiv G_0,\quad \bar{\bar{G}}_{m+} \equiv G,
\end{equation}
\end{subequations}
and, to simplify the notation, we will omit the dependence on $\omega$ of all Green tensors.

\subsection{Green tensor permutations} 
The Green tensor, that satisfies (\ref{rel:Green}), 
\begin{align}\label{rel:Green-11}
    \bar{\bar{G}}_{m+}(\mathbf{x}, \mathbf{y}) =& \bar{\bar{G}}_{0+}(\mathbf{x}, \mathbf{y}) +  \int_V \D^3z\,
    \beta(z) \bar{\bar{G}}_{0}(\mathbf{x}, \mathbf{z}) \bar{\bar{G}}(\mathbf{z}, \mathbf{x})),
\end{align}
and the reciprocity $\bar{\bar{G}}(\mathbf{x}, \mathbf{y}) = \bar{\bar{G}}^T(\mathbf{y}, \mathbf{x})$, have the following property:
\begin{align}
\bar{\bar{G}}(\mathbf{x}, \mathbf{y}) - \bar{\bar{G}}_{0}(\mathbf{x}, \mathbf{y}) &= \int_{V}\D^3z\,\beta(\mathbf{z})\bar{\bar{G}}_{0}(\mathbf{x}, \mathbf{z})\bar{\bar{G}}(\mathbf{z}, \mathbf{y}) \notag \\
&=  \int_{V}\D^3z\,\beta(\mathbf{z})\bar{\bar{G}}(\mathbf{x}, \mathbf{z})\bar{\bar{G}}_{0}(\mathbf{z}, \mathbf{y}). \label{prop:Green}
\end{align}
The first equality is just \eqref{rel:Green}.
The second equality can be obtained by taking the double transposed, using the reciprocity and finally using the first equality:
\begin{align}
& \int_{V}\D^3z\,\beta(\mathbf{z})\bar{\bar{G}}(\mathbf{x}, \mathbf{z})\bar{\bar{G}}_{0}(\mathbf{z}, \mathbf{y}) 
 \notag\\
& =  \left[  \int_{V}\D^3z\,\beta(\mathbf{z})\bar{\bar{G}}_{0}^T(\mathbf{z}, \mathbf{y})\bar{\bar{G}}^T(\mathbf{x}, \mathbf{z}) \right]^T\notag\\
& =  \left[  \int_{V}\D^3z\,\beta(\mathbf{z})\bar{\bar{G}}_{0}( \mathbf{y}, \mathbf{z})\bar{\bar{G}}(\mathbf{z}, \mathbf{x})\right]^T \notag\\
&= \left[  \bar{\bar{G}}( \mathbf{y}, \mathbf{x}) -  \bar{\bar{G}}_0( \mathbf{y}, \mathbf{x}) \right]^T  =  \bar{\bar{G}}( \mathbf{x}, \mathbf{y}) -  \bar{\bar{G}}_0( \mathbf{x}, \mathbf{y})  \notag\\
&=\int_{V}\D^3z\,\beta(\mathbf{z})\bar{\bar{G}}_{0}(\mathbf{x}, \mathbf{z})\bar{\bar{G}}(\mathbf{z}, \mathbf{y}).
\end{align}

{  
\subsection{Green tensor LDOS identity}

In this Appendix, we prove the Green tensor LDOS identity:
\begin{align}  \label{LDOS-identity-hrj}
    &\mathrm{Im}[\bar{\bar{G}}_{m+}(\mathbf{x}, \mathbf{y},\omega)] = \frac{\pi c^2}{2 \omega^3}
    \int d\kappa~  \mathbf{e}_\kappa(\mathbf{x})\otimes \mathbf{e}_\kappa^*(\mathbf{y}) ~\delta(\omega_\kappa-\omega) \notag \\
    &+ \frac{\omega^2}{c^2} \int \D^3 z\, \varepsilon_i(\mathbf{z}, \omega) \bar{\bar{G}}_{m+}(\mathbf{x},\mathbf{z}, \omega_{\kappa}) \bar{\bar{G}}_{m+}^*(\mathbf{z}, \mathbf{y}, \omega).
\end{align}
We will use \eqref{rel:Green} and the following relations:
\begin{align} \label{ImG0-Phi}
&\mathrm{Im}\, \bar{\bar{G}}_{0}(\mathbf{x}, \mathbf{y}, \omega)  =
\mathrm{Im}\, \bar{\bar{G}}_{0}^\perp(\mathbf{x}, \mathbf{y}, \omega) \notag\\
&= \frac{\pi c^2}{2 \omega} 
\int d\kappa~ \mathbf{\Phi}_\kappa(\mathbf{x}) \otimes \mathbf{\Phi}_{\kappa}(\mathbf{y}) \delta(\omega_\kappa-\omega),
\end{align}
where $\mathbf{\Phi}_{\kappa}$ are the transverse continuum eigenfunctions \eqref{def-Phi_kappa}  of $\nabla\times\nabla\times$, 
\begin{align}
\bar{\bar{G}}&(\mathbf{x}, \mathbf{z}_1) \left[ \mathbf{\Phi}(\mathbf{z}_1)\otimes  \mathbf{\Phi}(\mathbf{z}_2)\right] \bar{\bar{G}}^*(\mathbf{z}_2, \mathbf{y})  \notag \\
&=\left[\bar{\bar{G}}(\mathbf{x}, \mathbf{z}_1)  \mathbf{\Phi}(\mathbf{z}_1) \right] 
\otimes  \left[ \bar{\bar{G}}^{*T}(\mathbf{z}_2, \mathbf{y})  \mathbf{\Phi}(\mathbf{z}_2)\right] \notag \\
&=\left[\bar{\bar{G}}(\mathbf{x}, \mathbf{z}_1)  \mathbf{\Phi}(\mathbf{z}_1) \right] 
\otimes  \left[ \bar{\bar{G}}^{*}( \mathbf{y}, \mathbf{z}_2)  \mathbf{\Phi}(\mathbf{z}_2)\right],  \label{GGPP}
\end{align}
and \eqref{eqn:int_e_2}:
\begin{align}  \label{eqn:int_e_2-11}
\int_{V}   \D^3 z\,  \beta(\mathbf{z})  \bar{\bar{G}}(\mathbf{x}, \mathbf{z})  \mathbf{\Phi}_\kappa(\mathbf{z})  
 &=   \frac{1}{\omega_\kappa}\mathbf{e}_\kappa(\mathbf{x}) -\mathbf{\Phi}_\kappa(\mathbf{x}),   
\end{align} 
with $ \beta(\mathbf{z}) := \frac{\omega^2}{c^2}[\varepsilon( \mathbf{z}, \omega_\kappa) -1]$.
\\ \\
Consider the following difference
\begin{align}\label{start}
 D :=& \mathrm{Im}\, \bar{\bar{G}}(\mathbf{x}, \mathbf{y})
 - \int_{V}\D^3z\, \mathrm{Im}[\beta(\mathbf{z})]\bar{\bar{G}}(\mathbf{x}, \mathbf{z})\bar{\bar{G}}^*(\mathbf{z}, \mathbf{y}).
\end{align}
The imaginary part of (\ref{rel:Green}) is
\begin{align}\label{eqn:imGm}
\mathrm{Im}\, \bar{\bar{G}}(\mathbf{x}, \mathbf{y}) =& \mathrm{Im}\, \bar{\bar{G}}_{0}(\mathbf{x}, \mathbf{y}) \notag \\
&+ \frac{1}{2i}\left[\int_{V} \D^3 z\, \beta(\mathbf{z})\bar{\bar{G}}(\mathbf{x}, \mathbf{z})\bar{\bar{G}}_{0}(\mathbf{z}, \mathbf{y})\right. \notag \\
&-\left.  \int_{V} \D^3 z\, \beta^*(\mathbf{z})\bar{\bar{G}}^*_{0}(\mathbf{x}, \mathbf{z}) \bar{\bar{G}}^*(\mathbf{z}, \mathbf{y})\right],
\end{align}
where we used (\ref{prop:Green}) to rearrange the first integral.  Substituting $\mathrm{Im}[\beta(\mathbf{z})] = \frac{1}{2i}[\beta(\mathbf{z}) - \beta^*(\mathbf{z})]$ in the second term of (\ref{start}) and substituting (\ref{eqn:imGm}), we have

\begin{align}
D =& \mathrm{Im}\, \bar{\bar{G}}_{0}(\mathbf{x}, \mathbf{y}) \notag 
+ \frac{1}{2i}\Big[\int_{V} \D^3 z\, \beta(\mathbf{z})\bar{\bar{G}}(\mathbf{x}, \mathbf{z})\bar{\bar{G}}_{0}(\mathbf{z}, \mathbf{y})
\\
&- \int_{V} \D^3 z\, \beta^*(\mathbf{z})\bar{\bar{G}}^*_{0}(\mathbf{x}, \mathbf{z}) \bar{\bar{G}}^*(\mathbf{z}, \mathbf{y})\Big]  \notag\\
& -\frac{1}{2i}\Big[ \int_{V}\D^3z\, \beta(\mathbf{z})\bar{\bar{G}}(\mathbf{x}, \mathbf{z})\bar{\bar{G}}^*(\mathbf{z}, \mathbf{y}) \notag\\
&-\int_{V}\D^3z\, \beta^*(\mathbf{z})\bar{\bar{G}}(\mathbf{x}, \mathbf{z})\bar{\bar{G}}^*(\mathbf{z}, \mathbf{y}) 
 \Big]  \notag \\
=& \mathrm{Im}\, \bar{\bar{G}}_{0}(\mathbf{x}, \mathbf{y}) \notag \\
&+ \frac{1}{2i}\left\{\int_{V} \D^3 z\,\beta(\mathbf{z})\bar{\bar{G}}(\mathbf{x}, \mathbf{z})\left[\bar{\bar{G}}_{0}(\mathbf{z}, \mathbf{y}) - \bar{\bar{G}}^*(\mathbf{z}, \mathbf{y})\right]\right. \notag \\
&- \left. \int_{V} \D^3 z\,  \beta^*(\mathbf{z}) \left[\bar{\bar{G}}^*_{0}(\mathbf{x}, \mathbf{z}) - \bar{\bar{G}}(\mathbf{x}, \mathbf{z})\right] \bar{\bar{G}}^*(\mathbf{z}, \mathbf{y})\right\}. \label{D-prev}
\end{align}
Using 
 $\bar{\bar{G}}_{0}(\mathbf{x}, \mathbf{y}) - \bar{\bar{G}}^*_{0}(\mathbf{x}, \mathbf{y}) = 2i\, \mathrm{Im}\,\bar{\bar{G}}_{0}(\mathbf{x}, \mathbf{y})$ we can replace $ \bar{\bar{G}}_{0}$ by $ \bar{\bar{G}}_{0}^* + 2i\,\mathrm{Im}\,\bar{\bar{G}}_{0}$
  in the first integral and $ \bar{\bar{G}}_{0}^*$ by $ \bar{\bar{G}}_{0}- 2i \mathrm{Im}\,\bar{\bar{G}}_{0}$ in the second integral of \eqref{D-prev}
to write
 \begin{align}
D =& \mathrm{Im}\, \bar{\bar{G}}_{0}(\mathbf{x}, \mathbf{y}) \notag \\
&+ \frac{1}{2i} \int_{V} \D^3 z\,\beta(\mathbf{z})\bar{\bar{G}}(\mathbf{x}, \mathbf{z})\left[\bar{\bar{G}}_{0}^*(\mathbf{z}, \mathbf{y}) - \bar{\bar{G}}^*(\mathbf{z}, \mathbf{y})\right] \notag \\
&+ \int_{V} \D^3 z\,\beta(\mathbf{z})\bar{\bar{G}}(\mathbf{x}, \mathbf{z})
\left[  \mathrm{Im}\,\bar{\bar{G}}_{0}(\mathbf{z}, \mathbf{y})  \right]  \notag \\
&-  \frac{1}{2i}\int_{V} \D^3 z\,  \beta^*(\mathbf{z}) \left[\bar{\bar{G}}_{0}(\mathbf{x}, \mathbf{z}) - \bar{\bar{G}}(\mathbf{x}, \mathbf{z})\right] \bar{\bar{G}}^*(\mathbf{z}, \mathbf{y}) \notag  \\
&+ \int_{V} \D^3 z\,\beta^*(\mathbf{z})
\left[  \mathrm{Im}\,\bar{\bar{G}}_{0}(\mathbf{x}, \mathbf{z})   \right] \bar{\bar{G}}^*(\mathbf{z}, \mathbf{y}). 
\label{B16}
\end{align}
Now, using  (\ref{prop:Green}),
\begin{align}
\bar{\bar{G}}(\mathbf{x}, \mathbf{y}) - \bar{\bar{G}}_{0}(\mathbf{x}, \mathbf{y}) &= \int_{V}\D^3\mathrm{w}' \beta(\mathbf{w}')\bar{\bar{G}}_{0}(\mathbf{x}, \mathbf{w}')\bar{\bar{G}}(\mathbf{w}', \mathbf{y}) \notag \\
&=  \int_{V}\D^3z' \beta(\mathbf{z}')\bar{\bar{G}}(\mathbf{x}, \mathbf{z}')\bar{\bar{G}}_{0}(\mathbf{z'}, \mathbf{y}), &
\end{align}
 we  can replace $\bar{\bar{G}}_{0} -\bar{\bar{G}} $ in the first and in the third integrals  to obtain
\begin{align} 
\frac{1}{2i}&  \int_{V} \D^3 z\,  \beta~\bar{\bar{G}}  [ \bar{\bar{G}}_{0}^* - \bar{\bar{G}}^* ]  
- \frac{1}{2i} \int_{V} \D^3 \mathrm{w}\,  \beta^* \left[ \bar{\bar{G}}_{0} - \bar{\bar{G}} \right] \bar{\bar{G}}^* \notag \\
=&-\frac{1}{2i}  \int_{V} \D^3 z\,  \beta(\mathrm{z})\bar{\bar{G}}(\mathrm{x},\mathrm{z}) \notag \\
&\times \left[\int_{V}\D^3\mathrm{w}' \beta^*(\mathbf{w}')\bar{\bar{G}}^*_{0}(\mathbf{x}, \mathbf{w}')\bar{\bar{G}}^*(\mathbf{w}, \mathbf{y})  \right] \notag \\
&+ \frac{1}{2i} \int_{V} \D^3 \mathrm{w}\,  \beta^*(\mathrm{w})   \notag\\
&\times \left[\int_{V}\D^3\mathrm{z}' \beta(\mathbf{z}')\bar{\bar{G}}(\mathbf{x}, \mathbf{z}')\bar{\bar{G}}_{0}(\mathbf{z}', \mathbf{y}) \right] \bar{\bar{G}}^* (\mathrm{z}',\mathrm{y}) \notag \\
=& \int_{V}  \int_{V} \D^3 z\,\D^3 \mathrm{w}\,  \beta^*\beta~ \bar{\bar{G}}~ \frac{1}{2i} \left[ \bar{\bar{G}}_0- \bar{\bar{G}}_0^*  \right] \bar{\bar{G}}^*  \notag\\ 
=& \int_{V}  \int_{V} \D^3 z\,\D^3 \mathrm{w}\, \beta(\mathbf{z}) \beta^*(\mathbf{w}) \notag \\
&~~~~~~~~~ \times \bar{\bar{G}}(\mathbf{x}, \mathbf{z})
~\mathrm{Im}\,\bar{\bar{G}}_{0}(\mathbf{z},  \mathbf{w})~\bar{\bar{G}}^*(\mathbf{w}, \mathbf{y}).
\label{double-integral}
\end{align}
Substituting \eqref{ImG0-Phi},
 \begin{align}
\mathrm{Im}\, \bar{\bar{G}}_{0}(\mathbf{x}, \mathbf{y}, \omega) = \frac{\pi c^2}{2 \omega} 
\int d\kappa~ \mathbf{\Phi}_\kappa(\mathbf{x}) \otimes \mathbf{\Phi}_{\kappa}(\mathbf{y}),
\label{B19}
\end{align}
the double integral \eqref{double-integral} involves expressions of the form
\begin{align} 
& \int_{V}  \int_{V} \D^3 z\,\D^3 \mathrm{w}\, \beta(\mathbf{z}) \beta^*(\mathbf{w})~ \notag\\
&~~~~~~~~~ \times \bar{\bar{G}}(\mathbf{x}, \mathbf{z})   \left[ \mathbf{\Phi}(\mathbf{z})\otimes  \mathbf{\Phi}(\mathbf{w})\right] ~\bar{\bar{G}}^*(\mathbf{w}, \mathbf{y})  \notag \\
&=  \left[ \int_{V}   \D^3 z\,  \beta(\mathbf{z})  \bar{\bar{G}}(\mathbf{x}, \mathbf{z})  \mathbf{\Phi}(\mathbf{z})  \right] \notag\\
&~~~~~~~~~ \otimes 
 \left[  \int_{V}\D^3 \mathrm{w}\,  \beta^*(\mathbf{w}) \bar{\bar{G}}^*(\mathbf{y}, \mathbf{w})   \mathbf{\Phi}(\mathbf{w}) \right]
 \notag \\
 &=   \left[   \frac{\mathbf{e}_\kappa(\mathbf{x}) }{\omega_\kappa}-\mathbf{\Phi}_\kappa(\mathbf{x})    \right] 
 \otimes  \left[  \frac{\mathbf{e}_\kappa^*(\mathbf{y}) }{\omega_\kappa} -\mathbf{\Phi}_\kappa(\mathbf{y})    \right], 
 \label{B20}
\end{align} 
where we have used  \eqref{GGPP} and  \eqref{eqn:int_e_2-11}.

Similarly, the second and the fourth integral in \eqref{B16} involve expressions of the form
 \begin{align}
& \int_{V} \D^3 z\,\beta(\mathbf{z})\bar{\bar{G}}(\mathbf{x}, \mathbf{z})
\left[ \mathbf{\Phi}_\kappa(\mathbf{z}) \otimes \mathbf{\Phi}_{\kappa}(\mathbf{y})  \right]  \notag \\
&= \left[\int_{V} \D^3 z\,\beta(\mathbf{z})\bar{\bar{G}}(\mathbf{x}, \mathbf{z})
 \mathbf{\Phi}_\kappa(\mathbf{z})  \right] \otimes \mathbf{\Phi}_{\kappa}(\mathbf{y}) \notag\\
 &=  \left[\frac{\mathbf{e}_\kappa(\mathbf{x}) }{\omega_\kappa} -\mathbf{\Phi}_\kappa(\mathbf{x}) 
 \right] \otimes \mathbf{\Phi}_{\kappa}(\mathbf{y}) 
 \label{B21}
\end{align}
and
 \begin{align}
& \int_{V} \D^3 z\,\beta^*(\mathbf{z})
\left[ \mathbf{\Phi}_\kappa(\mathbf{x}) \otimes \mathbf{\Phi}_{\kappa}(\mathbf{z})   \right]
 \bar{\bar{G}}^*(\mathbf{z},\mathbf{y}) \notag\\
&= 
\mathbf{\Phi}_\kappa(\mathbf{x}) \otimes\left[ \int_{V} \D^3 z\,\beta^*(\mathbf{z})
   \bar{\bar{G}}^*(\mathbf{y},\mathbf{z})\mathbf{\Phi}_{\kappa}(\mathbf{z})   \right]  \notag\\
&=
\mathbf{\Phi}_\kappa(\mathbf{x}) \otimes  
    \left[ \frac{\mathbf{e}_\kappa^*(\mathbf{y}) }{\omega_\kappa} -\mathbf{\Phi}_\kappa(\mathbf{y}) 
 \right]. 
 \label{B22}
\end{align}
Putting together  \eqref{B16}, \eqref{double-integral}, \eqref{B19}, \eqref{B20}, and \eqref{B22}, we can write
\begin{align}
D =& \frac{\pi c^2}{2 \omega} 
\int d\kappa~ \delta(\omega_\kappa-\omega) \Bigg\{
\mathbf{\Phi}_\kappa(\mathbf{x}) \otimes \mathbf{\Phi}_{\kappa}(\mathbf{y}) \notag\\
&+  \left[ \frac{\mathbf{e}_\kappa(\mathbf{x}) }{\omega_\kappa} -\mathbf{\Phi}_\kappa(\mathbf{x}) 
 \right] \otimes \mathbf{\Phi}_{\kappa}(\mathbf{y}) \notag\\
& +
\mathbf{\Phi}_\kappa(\mathbf{x}) \otimes  
    \left[ \frac{\mathbf{e}_\kappa^*(\mathbf{y}) }{\omega_\kappa}-\mathbf{\Phi}_\kappa(\mathbf{y}) 
 \right]  \notag\\
&+ \left[  \frac{\mathbf{e}_\kappa(\mathbf{x}) }{\omega_\kappa} -\mathbf{\Phi}_\kappa(\mathbf{x})    \right] 
 \otimes  \left[  \frac{\mathbf{e}_\kappa^*(\mathbf{y}) }{\omega_\kappa} -\mathbf{\Phi}_\kappa(\mathbf{y})    \right]  
\Bigg\} \notag\\
&= \frac{\pi c^2}{2 \omega^3} 
\int d\kappa~ \delta(\omega_\kappa-\omega)~\mathbf{e}_\kappa(\mathbf{x})\otimes \mathbf{e}_\kappa^*(\mathbf{y}), 
\end{align}
which completes the proof.

}

%



\end{document}